\pdfoutput=1
\documentclass[11pt]{article}

\usepackage[a4paper,margin=2.5cm]{geometry}
\usepackage{amsmath,amssymb,amsfonts}
\usepackage{times}
\usepackage{microtype}
\usepackage{colortbl}
\usepackage{threeparttable}
\usepackage{booktabs}
\usepackage{placeins}
\usepackage{ragged2e}
\usepackage{graphicx}
\usepackage[font=small,labelfont=bf,labelsep=quad]{caption}
\usepackage[authoryear,round]{natbib}
\usepackage{authblk}
\usepackage[hidelinks]{hyperref}

\newcommand{\refrange}[3]{%
  \hyperref[#2]{#1~\ref*{#2}}--\hyperref[#3]{\ref*{#3}}}
\newcommand{\tto}{\mkern2mu{\to}\mkern2mu}
\newenvironment{acknowledgement}{\par\medskip\noindent\textbf{Acknowledgements}\quad}{\par}

\title{Estimating Treatment Effects in Complex Disease Dynamics: Immortal Time, Varying Hazards, and Multiple Time Scales}
\author{Yujun Xu}
\author{Ulrich Mansmann\thanks{Corresponding author: e-mail: \texttt{mansmann@ibe.med.uni-muenchen.de}}}
\affil{The Institute for Medical Information Processing, Biometry, and Epidemiology, Medical Faculty, Ludwig-Maximilians University, Marchioninistr.~15, 81377 Munich, Germany}
\date{}

\begin{document}
\raggedbottom

\maketitle

\begin{abstract}
In observational survival studies, delayed treatment initiation can induce immortal time bias in effect estimation when treatment groups are defined by future uptake. Despite originating from distinct predictive or causal inference paradigms, time-dependent approaches (\textit{e.g.}\ the time-dependent Cox model, landmark supermodel, multi-state model, and g-formula) are often treated as interchangeable in applied research when addressing immortal time bias. These models effectively reclassify person-time according to time-varying exposure, yet the interpretation of their target estimands depends on assumptions about the underlying disease process and clinically relevant time scales. We study effect estimation for allogeneic hematopoietic stem cell transplantation under baseline intention-to-transplant strategies in acute myeloid leukemia (\textit{i.e.}\ transplantation is planned for a prespecified time, subject to the patient's eligibility), within a multi-state framework allowing time-inhomogeneous Markov, semi-Markov, and state arrival extended semi-Markov dynamics. Using target trial emulation, we benchmark, on the probability scale, estimates derived from Cox-type time-dependent models and the g-formula against the corresponding theoretical values implied by the stochastic data-generating process. Decomposing the partial likelihood score, we show that single-time-scale models can be systematically biased when death intensities from different originating states are not proportional on a common time scale, as can occur when the clock is reset after intermediate events. We quantify these findings in a phase III simulation study and illustrate their implications in an application to data from three acute myeloid leukemia trials.
\end{abstract}


\section{Introduction}
\label{sec.1}
In many clinical settings, the treatment of interest following an initial therapy can be started only after a delay and only in patients who remain eligible until then, so that its receipt depends on the evolving patient history. This raises the question of how to quantify its population-level effect, relative to alternative treatments, in patients who are, once eligible, scheduled to receive it after a prespecified waiting time. A typical example is allogeneic hematopoietic stem cell transplantation (allo-HSCT) in acute myeloid leukemia (AML), which, where indicated, is given as consolidation after remission induction and can no longer be performed as planned if the patient becomes ineligible due to disease progression or dies beforehand \citep{dohner2022diagnosis}. Defining treatment groups by eventual receipt of allo-HSCT therefore introduces immortal time bias. For decision making at the time of eligibility (\textit{i.e.}\ at baseline), the relevant contrast is instead between strategies decided at that point, such as allo-HSCT planned for a prespecified time, irrespective of eventual receipt, versus chemotherapy only, in terms of the risk of the outcome within a given time horizon. The aim of this work is to examine whether, and under which assumptions about the disease process, commonly used approaches to immortal time bias (\textit{i.e.}\ the time-dependent Cox model, landmark supermodel, multi-state model, and g-formula) recover such contrasts on the probability scale when the process involves intermediate events and multiple clinically relevant time scales. Before formalizing this question and addressing it by simulation and an application to AML trial data, we review the issues that motivate it: immortal time bias and its usual remedies, assumptions about the disease process and its time scales implicit in these remedies, and the causal interpretation of hazard ratios in this setting.

In survival analysis, treatment effects are typically evaluated by contrasting intensity (hazard) functions between groups receiving different treatments, defined to start from a clinically relevant time origin (\textit{e.g.}\ disease diagnosis or treatment allocation). With a parametric or semi-parametric hazard model, the effects are often summarized by a hazard ratio comparing the groups' instantaneous event rates throughout, under a proportional hazards assumption. In randomized controlled trials, any delays in the initiation of treatment after allocation are accommodated under the intention-to-treat principle, which aligns with the treatment policy strategy in the estimand framework \citep{kahan2024estimands}, and randomization ensures baseline exchangeability between groups. However, in non-randomized studies, systematic differences in the timing of treatment initiation between groups due to reasons such as delayed donor availability or varying durations of preparative therapy can introduce immortal time bias \citep{suissa2008immortal, levesque2010problem}. The term `immortal time' refers to the period between the time origin and the initiation of the intended treatment, during which patients who eventually receive that treatment must, by design, remain event-free and may be offered the same standard of care as those in the control group. Therefore, this pre-treatment event-free period is a structural guarantee induced by defining treatment groups based on future initiation, rather than a protective effect of the treatment itself.

Immortal time bias has often been framed as a misclassification of person-time between exposure groups, which may reflect, at least in part, that `immortal time' is rhetorically striking and can inadvertently encourage a simplified interpretation as merely a classification problem. Under this view, converging evidence from systematic reviews of published literature, simulation studies, and mathematical derivations suggests that such misclassification systematically distorts estimates towards showing a beneficial effect when immortal time is incorrectly assigned to the treated group, often by a standard time-fixed Cox model with a baseline (ever/never) treatment indicator \citep{van2004time, austin2006quantifying, beyersmann2008easy}. This concept motivates the widespread use of Cox-type time-dependent (CTD) models, including the extended time-dependent Cox model and the landmark supermodel, among other related formulations, to reclassify person-time according to actual exposure status \citep{mantel1974evaluation, anderson1983analysis, van2011dynamic, gleiss2018unjustified, de2022analysis}. However, in applications of CTD models, assumptions about the hazards and the stochastic process that governs the evolving disease history and upon which the estimand targeted by these models depends are rarely made explicit.

From a process perspective, CTD methods are built on an instantaneous event rate defined relative to the current risk set and conditioned on the information available just before a given time point. The modeling step is then to separate the baseline component from a multiplicative covariate component, often within a single time scale (\textit{i.e.}\ the event process is treated as a single transition from the chosen origin to the outcome of interest), typically allowing the baseline hazard to change freely over time. Assuming the hazard at a given time not to depend on earlier history in any further way beyond what is captured by the covariate history yields a Markov-like description of the event process, conditional on the observed history (\textit{i.e.}\ the process is memoryless given the observed history). The process need not be time-homogeneous; indeed, time-inhomogeneity is clinically more plausible, as risks typically change over the course of the disease, reflecting the evolving clinical trajectory towards the absorbing state, which depends on the starting distribution of the process and on changes induced by intermediate events \citep{aalen2001understanding}. 

We note that events such as treatment initiation and disease progression are more than administrative time points: they represent boundary crossings of the underlying process, after which the governing transition dynamics may change. In the field of transplantation, for example, the hazard of death often surges immediately after transplantation and declines thereafter, so time since transplantation may be more prognostically informative than time since diagnosis and constitute an additional clinically relevant time scale. Such a pronounced peri-transplant spike, on the time scale since diagnosis, appears as an abrupt kink at each patient's time of transplantation, which a single common baseline hazard cannot adequately capture. As pointed out by \cite{iacobelli2013multiple}, whether to assume multiple time scales or a single time scale is a context-specific question, rather than a purely methodological choice. Nevertheless, analyzing event histories generated by such multi-state dynamics raises the question of whether CTD models that assume a single time scale and thus enforce a common baseline hazard shape from a shared origin, even after allowing for non-linear time effects via flexible covariate transformations, can represent additional time scales without distortion of regression parameters in time-inhomogeneous, non-Markovian settings. An alternative to representing additional time scales as time-dependent covariates is to project all time scales onto a single one \citep{farewell1979note}, on which the process is then Markovian. Such a scale is, however, an artificial construct and is valid only under a strong form of path independence \citep{oakes1995multiple}: the survival probability must depend only on the current values of the time scales it combines and not on the path by which they were reached; otherwise, single-time-scale models will conflate distinct post-transition dynamics.

Another issue in using the hazard ratio to infer causal treatment effects is that any effective treatment, whether beneficial or harmful, will inherently lead to systematic post-randomization imbalances in prognostic factors between those remaining at risk in the two groups over time, even under baseline comparability. This can be viewed as a form of continuous selection operating through the evolving risk sets \citep{hernan2010hazards, martinussen2022causality}; interpreting the hazard ratio as a contrast local in time $t$ is therefore problematic.   \citet{martinussen2020subtleties} define a causal hazard contrast at $t$ within the principal stratum of individuals who would survive up to $t$ under both treatments, but then acknowledge that membership in this subgroup is unobservable and that the contrast cannot be identified without untestable assumptions about the joint distribution of the potential survival times. \citet{beyersmann2025hazards} instead argue that hazard contrasts admit a causal interpretation when understood in a functional sense, that is, as contrasts between entire hazard functions rather than between hazards at a fixed time point, from which perspective the collider bias induced by conditioning on previous survival no longer arises.

Immortal time bias results from a related form of risk-set selection, which is, however, induced by the uptake process (who initiates treatment and when) rather than by treatment effects on survival. In observational settings, once treatment initiation is affected by disease prognosis and is not an external factor (\textit{i.e.}\ disease progression changes the likelihood of initiation, either deterministically or probabilistically), comparing treated and untreated person-time amounts to conditioning on a post-baseline event that is itself prognostically informative. Consequently, the hazard ratio of a time-dependent treatment indicator may reflect both treatment and selection effects through the risk sets, rather than a causal contrast between well-defined interventions. The functional argument of \citet{beyersmann2025hazards} concerns hazard functions under interventions decided at time zero; for time-dependent exposures, the authors themselves point to a multivariate hazards analysis to avoid immortal time bias. A functional reading of this hazard ratio would contrast the conditional hazards obtained by setting the indicator to 1 versus 0 for all individuals throughout follow-up, irrespective of eligibility. As treated person-time accrues only to individuals who have survived and remained eligible until treatment initiation, positivity is structurally violated, and identification requires an untestable extrapolation of the treatment effect observed after initiation to individuals who die or lose eligibility beforehand, as well as to the pre-treatment period. 

We agree with \citet{beyersmann2025hazards} and \citet{martinussen2020subtleties} that hazard-based contrasts should be routinely translated to the probability scale (\textit{e.g.}\ survival or cumulative event probability at a specific time point). The hazard ratio of a time-dependent treatment indicator, however, cannot be translated directly, as treatment status is an internal covariate whose path presupposes survival and eligibility; risk contrasts obtained by setting the treatment indicator to 1 versus 0 in CTD models inherit the problems above and generally admit a causal interpretation only if treatment is exogenous given the observed history \citep{cortese2010competing}. When treatment initiation depends on evolving disease prognosis, probability-scale estimands under well-defined interventions, such as treatment strategies defined at baseline, require explicit modeling of the joint treatment and covariate history, a step often left implicit in applied work.

Delayed treatment initiation, and hence immortal time, is inherent to many observational settings; a fundamental objective is therefore to determine how patients and physicians can be reliably advised on decision making at the start of the therapeutic process. In our motivating example, the additional survival benefit of allo-HSCT over standard chemotherapy remains unclear for subgroups of patients with AML \citep{gerstung2017precision,devillier2022depth,bornhauser2023allogeneic,niederwieser2024hematopoietic}, and randomized clinical trials for such questions are often infeasible due to accrual and logistical constraints; evidence is thus generated primarily from observational data. As outlined at the outset, we compare the cumulative probability of death by a time horizon $t^*$ (here, three years) under two strategies decided at baseline, namely, allo-HSCT scheduled at a prespecified time $\tau_{\mathrm{int}} < t^*$, with actual initiation subject to eligibility, versus chemotherapy only, and summarize this contrast by the risk ratio (RR), which we regard as important information for baseline decision making.

We note that this work is not intended to deliver an immediately deployable clinical decision tool, but to understand the potential biases of commonly used CTD models when applied to address immortal time bias, and to outline a path towards the future development of reliable tools. Although we recognize that existing statistical models were developed with distinct primary purposes, some geared towards effect estimation (\textit{e.g.}\ the time-dependent Cox model) and others towards prediction (\textit{e.g.}\ the landmark supermodel), they are frequently treated as interchangeable options for resolving immortal time bias in applied research \citep{gleiss2018unjustified, de2022analysis, devillier2022depth}. We also consider the g-formula, which attempts to account for the intrinsic dynamic covariates within a complex temporal process explicitly via the non-iterative conditional expectation algorithm \citep{hernan2020whatif,mcgrath2020gformula}. We discuss a formal setup by specifying a data-generating process to reflect clinical reality and utilizing a target trial emulation approach to generate intermediate-event histories and to obtain causally interpretable RR estimates \citep{matthews2022target}. That is to say, rather than advocating a single model, we articulate a coherent framework for defining the target estimand, comparing the estimators, and interpreting their estimates under realistic clinical dynamics in observational settings.

In \autoref{sec.2}, we formalize the motivating example as a multi-state process (\autoref{sec.2.1}), which serves as the basis for the data-generating process that produces patient histories under observational settings (\refrange{Sections}{sec.3.1}{sec.3.2}). We then define the causal estimand and its identification conditions and derive the true target quantity from this process (\autoref{sec.2.2}). Next, we consider different statistical approaches to account for immortal time bias (\autoref{sec.2.3}) and describe the target trial emulation used to estimate the RR under the two strategies (\autoref{sec.2.4}). In \autoref{sec.3}, we perform a simulation study, compare the empirical RRs derived from CTD models and the g-formula against their theoretical counterparts, and determine their bias. In \autoref{sec.4}, we explain the sources of the observed biases, and in \autoref{sec.5}, we discuss why treatment should be considered beyond a time-dependent covariate. We illustrate our findings using data from three AML trials of the German-Austrian AML Study Group (AMLSG) in \autoref{sec.aml}, before concluding in \autoref{sec.6}.

\section{Effect estimation from a stochastic process perspective}
\label{sec.2}

\subsection{Motivating example and formal framework} 
\label{sec.2.1}

Consider patients with intermediate-risk AML \citep{dohner2022diagnosis}: treatment typically begins with induction chemotherapy to achieve first complete remission (CR). Consolidation then proceeds with standard chemotherapy or allo-HSCT, although the survival benefit of allo-HSCT in this group remains unclear \citep{bornhauser2023allogeneic}. Referral to a transplant center and donor evaluation usually occur during the induction phase, and an intention-to-transplant strategy is typically defined once donor availability and early response are clarified; we refer to this moment as \textit{study entry} and consider a multi-state process thereafter. Under an intention-to-transplant strategy, allo-HSCT follows a preparation period and requires sustained disease control and adequate organ function, meaning that progression, complications, or death may occur before transplantation. The allo-HSCT strategy group therefore includes patients who receive allo-HSCT at the scheduled time and those who become ineligible beforehand. Accordingly, this dynamic process involves two intermediate events, \textit{allo-HSCT} and \textit{disease progression}, as well as \textit{death} as the outcome of interest; disease progression renders planned allo-HSCT infeasible. We acknowledge that this is a simplified abstraction that does not capture the full clinical complexity.

\begin{figure}[htbp]
    \centering
    \includegraphics[width=0.9\linewidth]{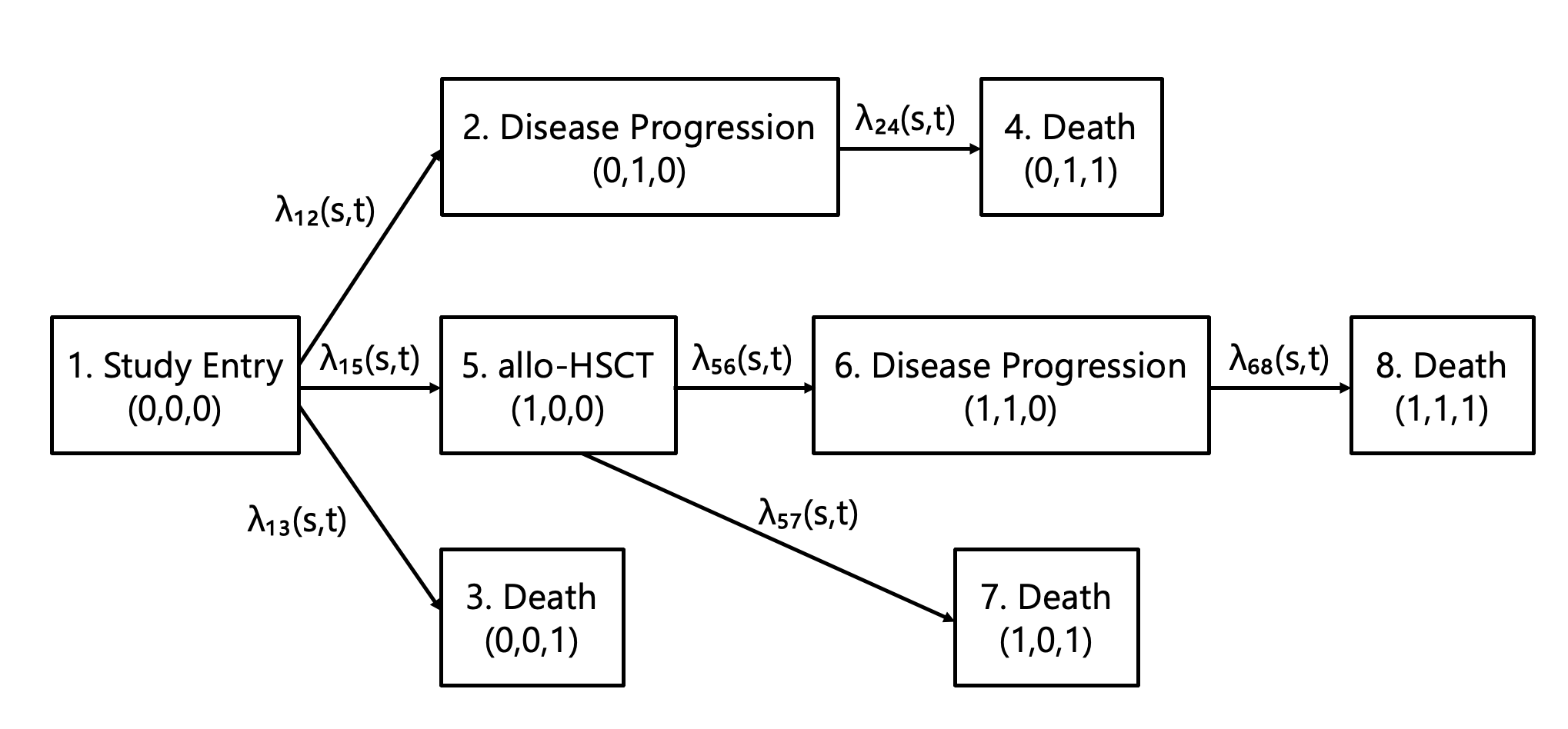}
    \caption{Disease trajectories considered in the multi-state framework}
    \label{fig_1}
\end{figure}

Let $\{X(t): t\ge 0\}$ denote the patient process after study entry, where $X(t)=(T_t,I_t,D_t)\in\mathcal{S}, \mathcal{S}:=\{0,1\}^3$, and $X(0)=(0,0,0)$. Throughout, we condition on a fixed baseline covariate vector $Z=z$ and suppress this conditioning in the notation. $T_t$, $I_t$, and $D_t$ indicate, respectively, whether allo-HSCT, disease progression, and death have occurred by time $t$; hence these components are non-decreasing in $t$: once equal to 1, they remain 1 thereafter. We define the corresponding occurrence times by $t_\mathrm{T}:=\inf\{t\ge 0: T_t=1\}$, 
$t_\mathrm{I}:=\inf\{t\ge 0: I_t=1\}$,
$t_\mathrm{D}:=\inf\{t\ge 0: D_t=1\}$,
with $\inf\emptyset:=\infty$ if the event never occurs. As depicted in \autoref{fig_1}, the process evolves in continuous time on the eight-state space
$\mathcal{S}=\{(0,0,0),\ldots,(1,1,1)\}$, and we index these states by $\{1,\ldots,8\}$ for notational convenience. States with $D_t=1$ are absorbing.

Different intensity models, defined by instantaneous transition intensities governing jumps between states at time $t$, characterize the process $\{X(t): t\ge 0\}$ at varying levels of complexity. The simplest model is based on time-homogeneous intensities $\lambda_{ij}$ for transitions from state $i$ to state $j$. Allowing the intensities to vary with time yields $\lambda_{ij}(t)$. More generally, we write \(\lambda_{ij}(s,t)\), where \(s\) denotes the time at which the current state \(i\) was entered, with \(s=0\) for \(i=1\). In the following, we use $\lambda_{ij}(s,t)$ as a unified notation, with $\lambda_{ij}$ and $\lambda_{ij}(t)$ corresponding to the special cases where $\lambda_{ij}(s,t)$ is constant in $(s,t)$ or depends only on $t$, respectively. 

The transition intensity matrix is therefore defined as \( Q(s,t) = (\lambda_{ij}(s,t): i,j \in \mathcal{S}) \). For $i\neq j$, $\lambda_{ij}(s,t)\ge 0$. Let $\lambda_i(s,t):=\sum_{j\neq i}\lambda_{ij}(s,t)$ denote the total exit rate from state $i$ at time $t$; the diagonal elements satisfy $\lambda_{ii}(s,t)=-\lambda_i(s,t)$. \( Q(s,t) \) can then be written as
\begin{equation}
\label{eq:intensity_matrix}
\scalebox{0.8}{$
\begin{array}{ccccccccc}
 & 000 & 010 & 001 & 011 & 100 & 110 & 101 & 111 \\
1. \;\;\;000 & -\lambda_{1}(s,t) & \lambda_{12}(s,t) & \lambda_{13}(s,t) & 0 & \lambda_{15}(s,t) & 0 & 0 & 0 \\
2. \;\;\;010 & 0 & -\lambda_{2}(s,t) & 0 & \lambda_{24}(s,t) & 0 & 0 & 0 & 0 \\
3. \;\;\;001 & 0 & 0 & 0 & 0 & 0 & 0 & 0 & 0 \\
4. \;\;\;011 & 0 & 0 & 0 & 0 & 0 & 0 & 0 & 0 \\
5. \;\;\;100 & 0 & 0 & 0 & 0 & -\lambda_{5}(s,t) & \lambda_{56}(s,t) & \lambda_{57}(s,t) & 0 \\
6. \;\;\;110 & 0 & 0 & 0 & 0 & 0 & -\lambda_{6}(s,t) & 0 & \lambda_{68}(s,t) \\
7. \;\;\;101 & 0 & 0 & 0 & 0 & 0 & 0 & 0 & 0 \\
8. \;\;\;111 & 0 & 0 & 0 & 0 & 0 & 0 & 0 & 0 \\
\end{array}
$}
\hphantom{0}\raisebox{-8 ex}{,}
\end{equation}
where the row names indicate the current state $i$ and the column names represent the potential next state $j$.

For a time-homogeneous Markov process with $Q(s,t)= Q$, the transition probability matrix from time $u$ to $v$ is given by the matrix exponential, \textit{i.e.}\ $P(u,v)=\exp\{(v-u)Q\}$ for $u<v$, where $P(u,v)_{ij}=\Pr\{X(v)=j \mid X(u)=i\}$. For a time-inhomogeneous Markov process with $Q(s,t) = Q(t)$, $P(u,v)$ is obtained by solving the Kolmogorov forward equation $\frac{\partial}{\partial t}P(u,t)=P(u,t)\,Q(t)$, $P(u,u)=I$, for $t\in[u,v]$. Moreover, it may be natural to assume a proportionality structure for transition intensities corresponding to the same clinical event across different histories; for example, as shown in \autoref{table_1}, $\lambda_{24}(s,t) = \delta\,\lambda_{13}(s,t)$, where $\delta>0$ is the hazard ratio for death after disease progression relative to death without disease progression.

We can now extend the time-inhomogeneous Markov process to two specific non-Markov settings: semi-Markov and state arrival extended semi-Markov processes \citep{putter2007tutorial}. In both cases, the transition intensities from the initial state, \textit{i.e.}\ $\{1\tto2, 1\tto3, 1\tto5\}$, coincide with those of the time-inhomogeneous Markov model. After an intermediate event (disease progression or allo-HSCT), the post-event intensities are modified by resetting the time scale, reflecting a change in the underlying disease dynamics. To formalize this, let $h_{ij}(\cdot)$ denote a specified function of time, such that $\lambda_{ij}(t)=h_{ij}(t)$ under the time-inhomogeneous Markov specification. In our setting, the history index \(s\) can be identified with the time of the most recent intermediate event prior to \(t\) (or \(s=0\) if no intermediate event has occurred); thus, for a given \(t\), we set $s=\max\big\{t_\mathrm{T}\,\mathbb{I}(t_\mathrm{T}<t),\ t_\mathrm{I}\,\mathbb{I}(t_\mathrm{I}<t),\ 0\big\}$, where \(\mathbb{I}(\cdot)\) is the indicator function. In the semi-Markov specification, for $i\neq j$, we model the transition intensities as functions of the elapsed time since the most recent intermediate event, $\lambda_{ij}(s,t)=h_{ij}\big(t-s\big)$, thereby introducing a reset time scale. In the state arrival extended semi-Markov specification, we further assume the time of disease progression to act as a prognostic factor for the process afterwards, by
replacing the hazard ratio $\delta$ for death after disease progression with a function $\rho(\cdot)\ge0$ of the arrival time, that is, $\lambda_{24}(s,t)=\rho(s)\,h_{13}(t-s)$ and $\lambda_{68}(s,t)=\rho(s)\,h_{57}(t-s)$. In the following, we focus on time-inhomogeneous Markov, semi-Markov, and state arrival extended semi-Markov processes.

\subsection{Causal estimand and theoretical treatment effects}
\label{sec.2.2}
As discussed earlier, to account for immortal time bias and enable a fair comparison, we contrast different baseline allo-HSCT strategies (defined by their scheduling times) with a chemotherapy-only strategy for mortality by a prespecified follow-up horizon \(t^*\). Formally, we define an intervention strategy \(I_{\mathrm{int}}\) that schedules allo-HSCT \textit{at} a prespecified time \(\tau_{\mathrm{int}}\), where \(\tau_{\mathrm{int}} \in (0,\infty]\), with \(\tau_{\mathrm{int}}=\infty\) indicating that no allo-HSCT is planned. We note that, in practice, the actual time of allo-HSCT may deviate from the planned time, depending on donor and recipient readiness. A clinically more realistic strategy would therefore initiate allo-HSCT \textit{within} a time window around $\tau_{\mathrm{int}}$, which would require integrating over the distribution of actual allo-HSCT times within that window. To keep the exposition focused on the identification and modeling issues, we restrict attention to allo-HSCT at the planned time $\tau_{\mathrm{int}}$; extension to window-based strategies is conceptually analogous. The actual time of allo-HSCT is then
\[
t_\mathrm{T}^{I_{\mathrm{int}}}
:=
\tau_{\mathrm{int}} \cdot \mathbb{I}\!\big\{\min(t_{\mathrm{I}},t_\mathrm{D})>\tau_{\mathrm{int}}\big\}
\;+\;
\infty \cdot \mathbb{I}\!\big\{\min(t_{\mathrm{I}},t_\mathrm{D})\le \tau_{\mathrm{int}}\big\},
\]
so that allo-HSCT occurs at \(\tau_{\mathrm{int}}\) if the patient remains alive and eligible up to \(\tau_{\mathrm{int}}\), and otherwise never occurs. The corresponding risk of death by time \(t^*\) under \(I_{\mathrm{int}}\) is 
$\Pr\!\big(t_\mathrm{D}^{I_{\mathrm{int}}}\le t^*\big)$. Hence, the population-level causal estimand can be formulated as
\[
E_{Z} \left\{ \Pr\!\big(t_\mathrm{D}^{I_{\mathrm{int}}}\le t^* \mid Z\big) \right\}
\quad \text{versus} \quad
E_{Z} \left\{ \Pr\!\big(t_\mathrm{D}^{I_\infty}\le t^* \mid Z\big) \right\},
\]
which contrasts the population-average probability of death up to \(t^*\) under strategy \(I_{\mathrm{int}}\) (\textit{i.e.}\ allo-HSCT planned at \(\tau_{\mathrm{int}}\)) with that under strategy \(I_\infty\) (\textit{i.e.}\ chemotherapy only), summarized by their ratio (RR), where \(Z\) denotes time-fixed covariates measured at baseline.

Identification of this estimand from observational data requires the following conditions, adapted from \citet{prosepe2025causal}: (i) \emph{consistency}: the potential time of death under $I_{\text{int}}$ equals the observed $t_{\mathrm{D}}$ for patients whose history is compatible with $I_{\text{int}}$, that is, who die or progress before $\tau_{\mathrm{int}}$ without allo-HSCT, or who receive allo-HSCT at $\tau_{\mathrm{int}}$ while alive and progression-free, and the potential time of death under $I_\infty$ equals the observed $t_{\mathrm{D}}$ for patients not transplanted during follow-up;  (ii) \emph{positivity}: among patients in state 1 at $\tau_{\mathrm{int}}$ (alive, progression-free, and not yet transplanted), the intensity of allo-HSCT at $\tau_{\mathrm{int}}$ is positive for all values of $Z$; (iii) \emph{conditional exchangeability}: among patients in state 1, initiation of allo-HSCT at any time is independent of the potential outcomes given $Z$, that is, it does not depend on unmeasured prognostic factors; (iv) \emph{independent censoring} given $Z$. We further assume that replacing the transition $1\tto5$ by allo-HSCT at $\tau_{\mathrm{int}}$ leaves the intensity functions of all other transitions unchanged, so that allo-HSCT affects the process only through entry into state 5 at $s=\tau_{\mathrm{int}}$; the risk $\Pr(t_{\mathrm{D}}^{I_{\text{int}}}\le t^*)$ is then given by the pathway decomposition below. \citet{prosepe2025causal} showed that this invariance follows from conditions analogous to (i)--(iv) in an illness--death model of treatment delay; our setting adds disease progression as a competing event that renders the scheduled treatment infeasible, an extension they identified as warranting further work. In the simulation study, (i)--(iv) and the invariance hold by construction; their plausibility in the application is discussed in \autoref{sec.aml}.

Given the transition intensity matrix $Q(s,t)$ in \autoref{sec.2.1}, we can calculate the theoretical risk of death by $t^*$ under strategy $I_{\mathrm{int}}$, $\Pr\!\big(t_\mathrm{D}^{I_{\mathrm{int}}}\le t^*\big)$. For notational simplicity, we write $\lambda_{ij}(t)$ and suppress the history index $s$ below, with $s=\tau_{\mathrm{int}}$ for the post-transplant intensities $\lambda_{56}$ and $\lambda_{57}$, and $s=r$, the time of progression, for the post-progression intensities $\lambda_{24}$ and $\lambda_{68}$; $\lambda_{ij}(t)$ is understood to denote the corresponding semi-Markov or state arrival extended semi-Markov intensities defined in \autoref{sec.2.1}. We also define two survival functions, before and after allo-HSCT, by 
\begin{gather*}
S_{1}(t) = \exp\left\{-\int_{0}^{t}\big(\lambda_{12}(u)+\lambda_{13}(u)\big)\,du\right\},\\
S_{5}(t) = \exp\left\{-\int_{\tau_{\mathrm{int}}}^{t}\big(\lambda_{56}(u)+\lambda_{57}(u)\big)\,du\right\}, \ \ t\ge \tau_{\mathrm{int}}.
\end{gather*}
For an intention-to-transplant strategy defined by a \textit{fixed} intervention time $\tau_{\mathrm{int}}<t^*$, $\Pr(t_\mathrm{D}^{I_{\mathrm{int}}}\le t^*)$ is evaluated by summing the probabilities of four mutually exclusive and exhaustive pathways. Specifically, $\Pr(t_\mathrm{D}^{I_{\mathrm{int}}}\le t^*) = P_1 + P_2 + P_3 + P_4$, where each component represents a distinct clinical trajectory:

\begin{itemize} 
\item[1.] Death occurs before disease progression and before $\tau_\mathrm{int}$ ($t_\mathrm{D} < t_\mathrm{I}$, $t_\mathrm{D} \le \tau_\mathrm{int}$):
\[
P_1 = 
\Pr(t_\mathrm{D}<t_\mathrm{I},\ t_\mathrm{D}\le \tau_{\mathrm{int}}) = \int_{0}^{\tau_{\mathrm{int}}} S_{1}(t)\,\lambda_{13}(t)\,dt.
\] 
\item[2.] Disease progression occurs before $\tau_\mathrm{int}$ and death occurs by $t^*$ ($t_\mathrm{I} < \tau_\mathrm{int}$, $t_\mathrm{I} < t_\mathrm{D} \le t^*$):
\[ 
P_2 = 
\Pr(t_\mathrm{I}<\tau_{\mathrm{int}},\ t_\mathrm{I}<t_\mathrm{D}\le t^*) = \int_{0}^{\tau_{\mathrm{int}}} S_{1}(r)\,\lambda_{12}(r)\, \Big[1-\exp\big\{-\int_{r}^{t^*}\lambda_{24}(u)\,du\big\}\Big]\,dr.
\] 
\item[3.] Allo-HSCT occurs exactly at $\tau_\mathrm{int}$ and death occurs by $t^*$ before any subsequent disease progression ($\tau_\mathrm{int} < t_\mathrm{D} < t_\mathrm{I}$, $t_\mathrm{D} \le t^*$):
\[ 
P_3 = 
\Pr(\tau_{\mathrm{int}} < t_\mathrm{D}<t_\mathrm{I}, t_\mathrm{D}\le t^*) = S_{1}(\tau_{\mathrm{int}}) \int_{\tau_{\mathrm{int}}}^{t^*} S_{5}(t)\,\lambda_{57}(t)\,dt.
\] 
\item[4.] Allo-HSCT occurs exactly at $\tau_\mathrm{int}$ and death occurs by $t^*$ after disease progression ($\tau_\mathrm{int} < t_\mathrm{I} < t_\mathrm{D} \le t^*$):
\[
P_4 = 
\Pr(\tau_{\mathrm{int}} <t_\mathrm{I}<t_\mathrm{D}\le t^*) = S_{1}(\tau_{\mathrm{int}}) \int_{\tau_{\mathrm{int}}}^{t^*} S_{5}(r)\,\lambda_{56}(r)\, \Big[1-\exp\big\{-\int_{r}^{t^*}\lambda_{68}(u)\,du\big\}\Big]\,dr. 
\] 
\end{itemize} 
Similarly, for the chemotherapy-only strategy with $\tau_{\mathrm{int}} = \infty$, the total risk of death reduces to the sum of two mutually exclusive pathways, $\Pr(t_\mathrm{D}^{I_\infty}\le t^*) = P_5 + P_6$:

\begin{itemize} 
\item[1.] Disease progression occurs first and death occurs by $t^*$ ($t_\mathrm{I} < t_\mathrm{D} \le t^*$):
\[ 
P_5 = 
\Pr(t_\mathrm{I}<t_\mathrm{D}\le t^*) = \int_{0}^{t^*} S_{1}(r)\,\lambda_{12}(r)\, \Big[1-\exp\big\{-\int_{r}^{t^*}\lambda_{24}(u)\,du\big\}\Big]\,dr.
\] 
\item[2.] Death occurs before disease progression and by $t^*$ ($t_\mathrm{D} < t_\mathrm{I}$, $t_\mathrm{D} \le t^*$):
\[ 
P_6=
\Pr(t_\mathrm{D}<t_\mathrm{I}, t_\mathrm{D}\le t^*) = \int_{0}^{t^*} S_{1}(t)\,\lambda_{13}(t)\,dt. 
\] 
\end{itemize}
The above integrals are evaluated numerically. See details in the \href{https://github.com/YX-IBE/stochastic_immortal_time_bias}{Online Repository}.

\subsection{Estimators for time-dependent treatment} \label{sec.2.3}

We investigate how four commonly used estimators for handling time-dependent covariates, namely, the time-dependent Cox model, landmark supermodel, multi-state model, and g-formula, capture the described dynamics in data generated under an assumed observational AML study.

A \textbf{time-dependent Cox model} partitions each individual's follow-up time at the observed occurrence times of allo-HSCT and disease progression, and codes $T(\cdot)$ and $I(\cdot)$ as piecewise-constant binary indicators for whether these events have occurred, updated at the start of each resulting interval. In a Markov scenario, the hazard function is
\[
h(t \mid Z, T(u), I(u)) = h_0(t)\exp\!\big\{\beta_{\mathrm{Z}}^\top Z + \beta_\mathrm{T} T(u) + \beta_\mathrm{I} I(u)\big\},
\]
where $h_0(t)$ denotes the baseline hazard, $Z$ is a vector of baseline covariates, and $u$ is the starting point of the interval containing $t$ (\textit{i.e.}\ the time at which the current state was entered). In semi-Markov and state arrival extended semi-Markov scenarios, the entry time $u$ of the current interval is included as an additional covariate \citep{iacobelli2013multiple} to capture its effect on the subsequent hazard at time $t$, through a linear or non-linear function $g(u)$:
\[
h(t \mid Z, u, T(u), I(u)) = h_0(t) \exp\big\{\beta_{\mathrm{Z}}^\top Z + \beta_\mathrm{T} T(u) + \beta_\mathrm{I} I(u) + g(u) \big\}.
\]

In the \textbf{landmark supermodel}, follow-up is partitioned by a grid of landmark time points $s$ with a small increment $\Delta s$. For each $s$, individuals who are event-free and under observation at $s$ contribute their follow-up to the window $(s, s+w]$, with $w>\Delta s$, yielding overlapping landmark datasets. Covariates are assessed at each landmark time point $s$, allowing for information updates and group reclassification \citep{van2008dynamic, van2011dynamic}. The resulting hazard model for $t \in (s, s+w]$ is
\[
h(t \mid Z, T(s), I(s), s, w)
= h_0(t \mid s, w)\exp\!\big\{\beta_{\mathrm{Z}}^\top Z + \beta_\mathrm{T}(s)\,T(s) + \beta_\mathrm{I}(s)\,I(s)\big\},
\]
where the baseline hazard in $(s, s+w]$ is allowed to vary with $s$ through
\[
h_0(t \mid s,w)= h_0(t \mid s_0,w)\exp\{g(s)\}, \text{with} \,\, g(s_0)=0.
\]
Additionally, the linearity assumption may be relaxed by modeling the coefficient functions $\beta_\mathrm{T}(s)$ and $\beta_\mathrm{I}(s)$, as well as the baseline adjustment $g(s)$, using functional forms (\textit{e.g.}\ quadratic or cubic polynomials). Multiple time scales are therefore accommodated through the landmark index $s$ and its flexible parameterization; no additional time-scale structure is introduced.

A \textbf{multi-state model} requires prespecification of disease transitions between states as in \autoref{fig_1} and fits a separate regression, \textit{e.g.}\ the Cox model, to each allowed transition $i\tto j$:
\[
\lambda_{ij}(t\mid Z)=\lambda_{ij,0}(t)\exp\{\gamma_{ij}^\top Z\},
\]
where $\lambda_{ij,0}(t)$ is the baseline hazard and $\gamma_{ij}$ the vector of regression coefficients for transition $i\tto j$ \citep{de2010mstate}. Let $\widehat{\Lambda}_{ij}(t)=\int_0^t \widehat{\lambda}_{ij}(u)\,du$ denote the estimated cumulative transition intensity and
$\widehat{\Lambda}(t)=(\widehat{\Lambda}_{ij}(t))_{i,j\in\mathcal{S}}$ the corresponding cumulative intensity matrix with
$\widehat{\Lambda}_{ii}(t)=-\sum_{j\neq i}\widehat{\Lambda}_{ij}(t)$.
The Aalen--Johansen product-integral estimator then yields the estimated transition probability matrix under the Markov assumption \citep{putter2007tutorial,de2010mstate} as
\[
\widehat{P}(u,v)=\prod_{(u,v]}\Bigl(I+d\widehat{\Lambda}(t)\Bigr), \quad u<v.
\]
As discussed previously, the multi-state framework also allows constraints across transitions by grouping transitions into strata that share a baseline hazard, while regression effects may be shared or allowed to differ. For instance, one may model all death transitions with a common baseline $\lambda_{\mathrm{D},0}(t)$ and common regression effects, with event history captured by the allo-HSCT indicator $T(t)$ and the disease progression indicator $I(t)$:
\[
\lambda_{ij}(t \mid Z,T(t),I(t))
=
\lambda_{\mathrm{D},0}(t)\exp\!\big\{\beta_\mathrm{Z}^\top Z+\beta_\mathrm{T}\,T(t)+\beta_\mathrm{I}\,I(t)\big\},
\quad (i\tto j)\in\mathcal A_{\mathrm{D}},
\]
where death transitions $\mathcal A_{\mathrm{D}}=\{1\tto 3,\,2\tto 4,\,5\tto 7,\,6\tto 8\}$. In semi-Markov and state arrival extended semi-Markov settings, multiple time scales can be incorporated by resetting the time origin after intermediate events (semi-Markov) or by augmenting subsequent transition models with state-entry times (state arrival extended semi-Markov), so that post-event transition intensities may depend on time since entry into the current state and/or on functions of state-entry times. However, since the post-event intensities then depend on the random times at which intermediate events occur, which are unknown at baseline, transition probabilities cannot be obtained directly from the Aalen--Johansen estimator and are typically computed by microsimulation, that is, by simulating trajectories of individual patients from the fitted model and averaging over a large number of them \citep{krijkamp2018microsimulation}.

The \textbf{g-formula} operates on discrete time intervals and performs standardization for risk estimation across all theoretical trajectories \citep{hernan2020whatif,mcgrath2020gformula}. Follow-up is partitioned into $K$ intervals $[t_{k-1}, t_k)$, $k = 1, \ldots, K$, with $t_0 = 0$ and $t_K = t^*$; the intervals are sufficiently short that event rates within an interval are low. Let $T_k$, $I_k$, and $D_k$ indicate whether allo-HSCT, disease progression, and death have occurred by the end of interval $k$, respectively. Since event times are recorded in continuous time and then discretized for the g-formula, two or more events may fall within the same interval. In such cases, events are assumed to occur in the order allo-HSCT, disease progression, and death. The event processes are modeled via pooled logistic regressions for the discrete-time hazards, conditional on the observed history and on being at risk at the start of interval $k$, that is, alive and free of the respective event and, for allo-HSCT, also progression-free. For interval $[t_{k-1},t_k)$, for example, we may specify
\begin{align*}
\operatorname{logit}\Pr\bigl(T_k=1 \mid \mathcal{H}_{k-1}\bigr)
&= \alpha_{\mathrm{T},0} + \alpha_{\mathrm{T},1}\,t_{k-1},\\
\operatorname{logit}\Pr\bigl(I_k=1 \mid \mathcal{H}_{k-1}\bigr)
&= \alpha_{\mathrm{I},0} + \alpha_{\mathrm{I},1}\,t_{k-1} + \phi_\mathrm{IZ}^\top Z + \phi_\mathrm{IT}^\top f_{\mathrm{T}}(\bar T_{k-1}),\\
\operatorname{logit}\Pr\bigl(D_k=1 \mid \mathcal{H}_{k-1}\bigr)
&= \alpha_{\mathrm{D},0} + \alpha_{\mathrm{D},1}\,t_{k-1} + \phi_{\mathrm{DZ}}^\top Z + \phi_{\mathrm{DT}}^\top f_{\mathrm{T}}(\bar T_{k-1}) + \phi_{\mathrm{DI}}^\top f_{\mathrm{I}}(\bar I_{k-1}),
\end{align*}
where $\mathcal{H}_{k-1}$ denotes the observed history at $t_{k-1}$, including the baseline covariates $Z$ and the event histories $\bar T_{k-1}=(T_0,\ldots,T_{k-1})$, $\bar I_{k-1}=(I_0,\ldots,I_{k-1})$, and $\bar D_{k-1}=(D_0,\ldots,D_{k-1})$, with $T_0=I_0=D_0=0$. To summarize these histories, one could use
\[
f_{\mathrm{T}}(\bar T_{k-1}) = \Bigl(T_{k-1},\; \frac{1}{k}\sum_{l=0}^{k-1} T_l\Bigr),
\qquad
f_{\mathrm{I}}(\bar I_{k-1}) = \Bigl(I_{k-1},\; \frac{1}{k}\sum_{l=0}^{k-1} I_l\Bigr).
\] 
The first component of $f_{\mathrm{T}}$ indicates the current allo-HSCT status and the second the proportion of elapsed follow-up spent after allo-HSCT, which equals $0$ before allo-HSCT and approximately $1-t_{\mathrm{T}}/t_{k-1}$ thereafter. The summary $f_{\mathrm{I}}$ is defined analogously for disease progression. Importantly, the g-formula introduces no additional time structure by default; post-event dynamics are represented only through covariates constructed from $\mathcal{H}_{k-1}$.

\subsection{Target trial emulation} 
\label{sec.2.4}
Estimating the empirical RR contrasting two therapeutic strategies with the estimators discussed in \autoref{sec.2.3} requires the intermediate-event history, from which the probability of the outcome is computed. However, both allo-HSCT and disease progression are endogenous events on the disease trajectory: they lie on the causal pathway to death and, in observational data, are subject to competing-event truncation, so their time of occurrence may be undefined.

The g-formula addresses this by adopting a built-in target trial emulation approach through Monte Carlo standardization: parameters from the pooled logistic models (\autoref{sec.2.3}) are used to simulate a sufficiently large number of hypothetical patient trajectories forward in discretized time, from study entry up to $t^*$, under the defined therapeutic strategies. 
Event indicators are updated sequentially at each time step, conditional on the history and on being at risk at the start of the interval, thereby allowing the corresponding risk at $t^*$ to be computed.

The time-dependent Cox model and the landmark supermodel do not, by themselves, capture the joint dynamics of allo-HSCT and disease progression and thus cannot determine whether a patient remains eligible for allo-HSCT at $\tau_{\mathrm{int}}$ under a given strategy. A fitted multi-state model, by contrast, describes these joint dynamics explicitly and can be used to simulate counterfactual intermediate-event histories under the defined strategies by microsimulation. Under the identification conditions of \autoref{sec.2.2}, we therefore implement the following target trial emulation procedure:

\begin{enumerate}
\item Given an intervention strategy $I_{\mathrm{int}}$ with scheduled allo-HSCT time $\tau_{\mathrm{int}}$, construct a working distribution for the \textit{counterfactual} disease progression time under $I_{\mathrm{int}}$, $F_{\mathrm{I}}(t\mid Z, I_{\mathrm{int}})$, using the estimated transition-specific hazards from the multi-state model. Let $\Lambda_{12}(t\mid Z)$ and $\Lambda_{56}(t\mid Z)$ denote the cumulative transition hazards for disease progression without prior allo-HSCT ($1\tto 2$) and with allo-HSCT ($5\tto 6$), given baseline covariates $Z$. A piecewise specification is then
\[
F_{\mathrm{I}}(t\mid Z, I_{\mathrm{int}})=
\begin{cases}
1-\exp \bigl\{-\Lambda_{12}(t\mid Z)\bigr\}, & t\le \tau_{\mathrm{int}},\\[4pt]
1-\exp \bigl\{-\Lambda_{56}(t\mid Z)+\Lambda_{56}(\tau_{\mathrm{int}}\mid Z)-\Lambda_{12}(\tau_{\mathrm{int}}\mid Z)\bigr\}, & t>\tau_{\mathrm{int}}.
\end{cases}
\]
In particular, under no planned allo-HSCT, we write $\tau_{\mathrm{int}}=\infty$, so that
\[
F_{\mathrm{I}}(t\mid Z, I_\infty)=1-\exp \bigl\{-\Lambda_{12}(t\mid Z)\bigr\}.
\]
Under the semi-Markov and state arrival extended semi-Markov specifications, the post-transplant cumulative hazard $\Lambda_{56}$ is evaluated on the time since allo-HSCT accordingly (see the \href{https://github.com/YX-IBE/stochastic_immortal_time_bias}{Online Repository}).
\item Use the inverse transform sampling method to simulate the counterfactual disease progression time $t_\mathrm{I}^{I_{\mathrm{int}}}$ for a sufficiently large number of hypothetical patients from $F_{\mathrm{I}}(\cdot\mid Z, I_{\mathrm{int}})$, and define the event indicator for disease progression by $t^*$, $I^{I_{\mathrm{int}}}$, as
\[
I^{I_{\mathrm{int}}}
=
\mathbb{I}\!\left(t^* \ge t_\mathrm{I}^{I_{\mathrm{int}}}\right).
\]

\item Define the time of allo-HSCT and the event indicator for allo-HSCT by $t^*$, $T^{I_{\mathrm{int}}}$, as
\[
t_{\mathrm{T}}^{I_{\mathrm{int}}}
=
\tau_{\mathrm{int}}\,
\mathbb{I}\!\left(t_{\mathrm{I}}^{I_{\mathrm{int}}} > \tau_{\mathrm{int}}\right)
\;+\;
\infty\,
\mathbb{I}\!\left(t_{\mathrm{I}}^{I_{\mathrm{int}}} \le \tau_{\mathrm{int}}\right),
\]
\[
T^{I_{\mathrm{int}}}
=
\mathbb{I}\!\left(t^* \ge \tau_{\mathrm{int}}\right)\,
\mathbb{I}\!\left(t_\mathrm{I}^{I_{\mathrm{int}}} > \tau_{\mathrm{int}}\right).
\]
\end{enumerate}
The resulting counterfactual history of allo-HSCT and disease progression can then be used as input for the time-dependent Cox model, landmark supermodel, and multi-state model for subsequent estimation of the risk of death by $t^*$. Finally, empirical effect estimates from these estimators can be compared with the theoretical effects in \autoref{sec.2.2} to assess bias.

\section{Simulation study} 
\label{sec.3}
To evaluate the four estimators considered in \autoref{sec.2.3}, we conducted a phase III simulation study \citep{heinze2024phases} comprising 1{,}000 replicate observational datasets, each with 1{,}000 AML patients generated according to the multi-state process $\{X(t):t\ge 0\}$ described in \autoref{sec.2.1}. In the envisaged observational study, patients were monitored monthly for three years after study entry, and allo-HSCT, disease progression, and death status were updated at each visit, while event times were recorded in continuous time. In addition, baseline heterogeneity was introduced via two binary covariates $Z_1$ (affecting disease progression and death) and $Z_2$ (affecting death only), each combination of which defined a subgroup of 250 patients. For each observational dataset, we performed target trial emulation and derived estimates from each estimator.

\subsection{Simulation parameters}
\label{sec.3.1}

\begin{table}[!ht]
\centering
\begin{minipage}{0.57\linewidth}
\centering
\caption{Transition intensity specifications across scenarios with a Weibull death hazard}
\label{table_1}

\begin{threeparttable}
\scriptsize 
\renewcommand{\arraystretch}{1.15}
\setlength{\tabcolsep}{4pt} 

\begin{tabular}{c c c c}
\toprule
& \textbf{Markov} & \textbf{Semi-Markov} &
  \textbf{\begin{tabular}[c]{@{}c@{}}State arrival extended\\ semi-Markov\end{tabular}} \\
\midrule
\multicolumn{4}{c}{\textbf{No treatment effect}}\\
\midrule
$\lambda_{12}(\cdot)$ & $h_{12}(t)$ & $h_{12}(t)$ & $h_{12}(t)$ \\
$\lambda_{13}(\cdot)$ & $h_{13}(t)$ & $h_{13}(t)$ & $h_{13}(t)$ \\
$\lambda_{15}(\cdot)$ & $h_{15}(t)$ & $h_{15}(t)$ & $h_{15}(t)$ \\
$\lambda_{24}(\cdot)$ & $2\cdot h_{13}(t)$ & $2\cdot h_{13}(t-s)$ & $\rho(s)\cdot h_{13}(t-s)$ \\
$\lambda_{56}(\cdot)$ & $h_{12}(t)$ & $h_{12}(t-s)$ & $h_{12}(t-s)$ \\
$\lambda_{57}(\cdot)$ & $h_{13}(t)$ & $h_{13}(t-s)$ & $h_{13}(t-s)$ \\
$\lambda_{68}(\cdot)$ & $2\cdot h_{13}(t)$ & $2\cdot h_{13}(t-s)$ & $\rho(s)\cdot h_{13}(t-s)$ \\
\midrule
\multicolumn{4}{c}{\textbf{Treatment effect on death}}\\
\midrule
$\lambda_{12}(\cdot)$ & $h_{12}(t)$ & $h_{12}(t)$ & $h_{12}(t)$ \\
$\lambda_{13}(\cdot)$ & $h_{13}(t)$ & $h_{13}(t)$ & $h_{13}(t)$ \\
$\lambda_{15}(\cdot)$ & $h_{15}(t)$ & $h_{15}(t)$ & $h_{15}(t)$ \\
$\lambda_{24}(\cdot)$ & $2\cdot h_{13}(t)$ & $2\cdot h_{13}(t-s)$ & $\rho(s)\cdot h_{13}(t-s)$ \\
$\lambda_{56}(\cdot)$ & $h_{12}(t)$ & $h_{12}(t-s)$ & $h_{12}(t-s)$ \\
$\lambda_{57}(\cdot)$ & $0.5\cdot h_{13}(t)$ & $0.5\cdot h_{13}(t-s)$ & $0.5\cdot h_{13}(t-s)$ \\
$\lambda_{68}(\cdot)$ & $2\cdot 0.5\cdot h_{13}(t)$ & $2\cdot 0.5\cdot h_{13}(t-s)$ & $\rho(s)\cdot 0.5\cdot h_{13}(t-s)$ \\
\midrule
\multicolumn{4}{c}{\textbf{Treatment effect on disease progression and death}}\\
\midrule
$\lambda_{12}(\cdot)$ & $h_{12}(t)$ & $h_{12}(t)$ & $h_{12}(t)$ \\
$\lambda_{13}(\cdot)$ & $h_{13}(t)$ & $h_{13}(t)$ & $h_{13}(t)$ \\
$\lambda_{15}(\cdot)$ & $h_{15}(t)$ & $h_{15}(t)$ & $h_{15}(t)$ \\
$\lambda_{24}(\cdot)$ & $2\cdot h_{13}(t)$ & $2\cdot h_{13}(t-s)$ & $\rho(s)\cdot h_{13}(t-s)$ \\
$\lambda_{56}(\cdot)$ & $0.5\cdot h_{12}(t)$ & $0.5\cdot h_{12}(t-s)$ & $0.5\cdot h_{12}(t-s)$ \\
$\lambda_{57}(\cdot)$ & $0.5\cdot h_{13}(t)$ & $0.5\cdot h_{13}(t-s)$ & $0.5\cdot h_{13}(t-s)$ \\
$\lambda_{68}(\cdot)$ & $2\cdot 0.5\cdot h_{13}(t)$ & $2\cdot 0.5\cdot h_{13}(t-s)$ & $\rho(s)\cdot 0.5\cdot h_{13}(t-s)$ \\
\bottomrule
\end{tabular}

\vspace{4pt}
\renewcommand{\TPTnoteSettings}{\scriptsize\raggedright}
\begin{tablenotes}[flushleft]
\item[1] \(h_{12}(t)=0.12\cdot0.55\cdot t^{0.55-1},\;
h_{13}(t)=0.048\cdot0.60\cdot t^{0.60-1}\),
and \(h_{15}(t)=0.012\cdot1.5\cdot t^{1.5-1}.\)
\item[2] \(s\) is the most recent intermediate-event time before \(t\), or \(0\) if neither has occurred.
\item[3] \(\rho(s)=1+\dfrac{1}{1+\exp(s-6)}.\)
\item[4] Two baseline binary prognostic factors define four equally sized subgroups, with groups 1--4 corresponding to $(Z_1,Z_2)=(0,0),(1,0),(0,1),(1,1)$; group 1 is shown. Weibull scale parameter for $h_{12}$: 0.12, 0.16, 0.12, 0.16; for $h_{13}$: 0.048, 0.060, 0.072, 0.084.
\item[5] \(t\) and \(s\) are measured in months since study entry.
\end{tablenotes}

\end{threeparttable}
\end{minipage}
\end{table}

We considered nine scenarios (\autoref{table_1}) obtained by crossing three process specifications (time-inhomogeneous Markov, semi-Markov, and state arrival extended semi-Markov) with three effect settings (no treatment effect, effect on death, and effect on both disease progression and death). The intensities of disease progression and death were assumed to decline over time, reflecting a higher-risk period in early follow-up, followed by declining risks as patients stabilize and tolerate therapy, whereas the intensity of allo-HSCT was assumed to increase over time as donor search and transplant preparation progress. Disease progression multiplies the death intensity by a factor of $\delta=2$ or, in the state arrival extended semi-Markov setting, by a factor $\rho(s)$ that decreases with a longer progression-free sojourn time. Transition intensities corresponding to the same clinical event were grouped into strata with proportionality constraints within each stratum, \textit{e.g.}\ $\lambda_{24}(\cdot)=\delta\,\lambda_{13}(\cdot)$, and the effects of disease progression and allo-HSCT on the death intensity were assumed not to interact on the hazard scale. Non-Markov behavior was induced by resetting the time scale ($t-s$) after intermediate events in the semi-Markov scenario and, in the state arrival extended semi-Markov scenario, in addition by allowing the death intensities after progression to depend on the arrival time through $\rho(s)$. In the no-effect setting, the post-transplant intensities share the functional form of their pre-transplant counterparts (\textit{e.g.}\ $h_{13}$ for death), that is, allo-HSCT has no multiplicative effect; under clock reset, however, allo-HSCT still resets the time scale, so that these intensities depend on the time since allo-HSCT rather than on the time since study entry, \textit{e.g.}\ $\lambda_{57}(s,t)=h_{13}(t-s)$. In the two settings with treatment effect, allo-HSCT additionally halves the post-transplant death intensity, or both the death and progression intensities, respectively.

Here we present results with transitions specified by Weibull hazards. In addition, parameters for simulation under log-logistic death hazards and the corresponding results are reported in \hyperref[app.b]{Appendix~B} (\refrange{Tables}{table_b1}{table_b2}, \refrange{Figures}{fig_b1}{fig_b3}) to explore the behavior of the estimators under alternative hazard shapes for methodological illustration.

\subsection{Data-generating mechanism} 
\label{sec.3.2}

Data generation was performed using a modified Gillespie algorithm to simulate continuous event times \citep{gillespie1976general,gillespie1977exact,gillespie2007stochastic}. 
As in \autoref{sec.2.1}, suppose the process has entered state $i\in\mathcal{S}$ at time $u$, so that the history index is $s=u$. The simulation iterates the following steps:

\begin{enumerate}
\item Given $X(u)=i$, simulate the next jump time $v>u$ using the conditional cumulative distribution function
\[
F_{i,u}(v)=1-\exp\big\{-\int_u^{v}\lambda_i\big(u,r\big)\,dr\big\}.
\]

\item Conditional on a jump occurring at time $v$ from state $i$, sample the next state $j\neq i$ from
\[
\Pr\{X(v)=j \mid X(u)=i,\text{jump at }v\}
=
\frac{\lambda_{ij}\big(u,v\big)}{\sum_{q\neq i}\lambda_{iq}\big(u,v\big)}.
\]

\item Update $(u,i)$ with $(v,j)$, and repeat Steps 1--2 until an absorbing state is reached or follow-up ends.

\item Repeat Steps~1--3 independently for each individual and for each simulated dataset.
\end{enumerate}
In semi-Markov and state arrival extended semi-Markov scenarios, clock reset is handled implicitly through the dependence of $\lambda_{ij}(s,t)$ on $t-s$; hence no additional time shift of the sampling distribution is required.

\subsection{Software}
This study was conducted using R version 4.3.2. A detailed R script containing the simulation configurations is available in the \href{https://github.com/YX-IBE/stochastic_immortal_time_bias}{Online Repository}.

\subsection{Simulation results}

\begin{figure}[!htbp]
    \centering
    \captionsetup{font=small}
    \includegraphics[width=\linewidth]
    {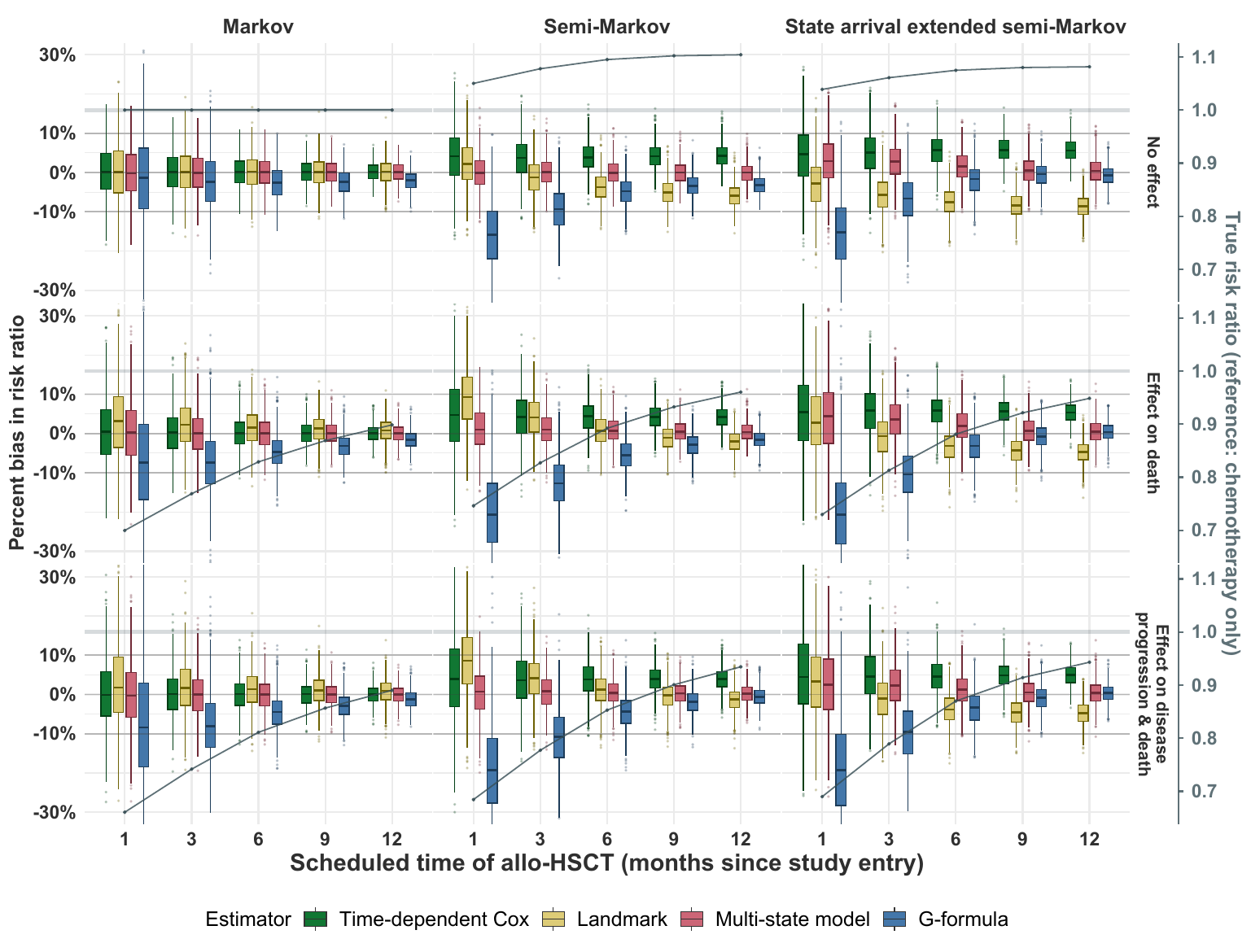}
    \caption[Bias in risk ratio for 3-year mortality (Weibull distribution)]{Main simulation results under Weibull death transition intensities: percent bias of the four estimators in the estimated 3-year mortality risk ratio (RR; reference: chemotherapy only). The x-axis indexes intervention strategies \(I_{\mathrm{int}}\) by planned allo-HSCT time \(\tau_{\mathrm{int}}\) (months since study entry), with one box plot per \(\tau_{\mathrm{int}}\); the y-axis is \(\{(\widehat{\mathrm{RR}}-\mathrm{RR}_{\mathrm{true}})/\mathrm{RR}_{\mathrm{true}}\}\times 100\%\). The time-dependent Cox model includes linear terms in the times of allo-HSCT and disease progression in the semi-Markov and state arrival extended semi-Markov settings. The landmark supermodel uses \(\Delta s=1\), \(w=6\), quadratic \(\beta(s)\), and cubic \(g(s)\). The g-formula uses the pooled logistic models of \autoref{sec.2.3} for disease
    progression and death, with a linear term in time, under all three specifications. Solid curves show \(\mathrm{RR}_{\mathrm{true}}\) from the transition intensity matrix. The light horizontal band marks $\mathrm{RR}_{\mathrm{true}} = 1$ (right axis). 
    In the no-effect setting, allo-HSCT has no multiplicative effect on the transition intensities (\autoref{table_1}); under the semi-Markov and state arrival extended semi-Markov specifications, however, post-transplant intensities restart on the time since allo-HSCT, \textit{e.g.}\ $h_{13}(t-s)$, which induces a transient excess hazard and hence $\mathrm{RR}_{\mathrm{true}}>1$.}
    \label{fig_2}
\end{figure}

In \autoref{fig_2}, we present results for the nine scenarios. Columns correspond to time-inhomogeneous Markov, semi-Markov, and state arrival extended semi-Markov processes, and rows to the three effect settings (no effect, effect on death, and effect on both disease progression and death). Each panel summarizes 1{,}000 simulation replicates, with each replicate comprising an observational cohort of 1{,}000 AML patients.

For each scenario, we contrast the 3-year mortality risk under strategy \(I_{\mathrm{int}}\), in which allo-HSCT is scheduled at \(\tau_{\mathrm{int}} \in \{1,3,6,9,12\}\) months after study entry, with that under the chemotherapy-only strategy \(I_{\infty}\), summarized by the RR. The dark solid curves show the theoretical RR derived from the transition intensity matrix (\autoref{sec.2.2}). If allo-HSCT confers no survival benefit, we expect RR$=$1 across strategies under the time-inhomogeneous Markov process (top-left panel). When allo-HSCT is beneficial, the magnitude of the true survival benefit attenuates as \(\tau_{\mathrm{int}}\) is delayed, since more patients experience disease progression or die before the scheduled allo-HSCT and therefore can no longer receive it. Under the semi-Markov and state arrival extended semi-Markov processes, however, a transient peri-transplant excess hazard can induce RR$>$1 even without a multiplicative effect of allo-HSCT (top-middle/-right panels). Since the death intensity declines over time, restarting it at an early allo-HSCT changes it little, as the background intensity is still high, whereas restarting it later raises it well above the already lower background intensity; the excess cumulative hazard up to $t^*$ is therefore smaller for earlier $\tau_{\mathrm{int}}$, which translates into a relatively smaller RR (middle/right columns) and reinforces the attenuation described above when allo-HSCT is beneficial.

\begin{table}[!htbp]
\renewcommand{\arraystretch}{1.20}
\setlength{\tabcolsep}{3.5pt}
\centering
\caption[Simulation performance measures]{Performance measures of risk ratio estimates for 3-year mortality across simulation scenarios under Weibull death transition intensities at $\tau_{\mathrm{int}}$ = 3 months}
\label{table_2}

\medskip
\noindent (A) Bias, empirical standard error (EmpSE), and mean squared error (MSE)

\resizebox{\textwidth}{!}{%
\begin{tabular}{l
                *{4}{>{\centering\arraybackslash}m{1.25cm}
                      >{\centering\arraybackslash}m{1.25cm}
                      >{\centering\arraybackslash}m{1.25cm}}}
\toprule
& \multicolumn{3}{c}{\textbf{Time-dependent Cox}}
& \multicolumn{3}{c}{\textbf{Landmark}}
& \multicolumn{3}{c}{\textbf{Multi-state model}}
& \multicolumn{3}{c}{\textbf{G-formula}} \\
\cmidrule(lr){2-4}\cmidrule(lr){5-7}\cmidrule(lr){8-10}\cmidrule(lr){11-13}
& \textbf{Bias} & \textbf{EmpSE} & \textbf{MSE}
& \textbf{Bias} & \textbf{EmpSE} & \textbf{MSE}
& \textbf{Bias} & \textbf{EmpSE} & \textbf{MSE}
& \textbf{Bias} & \textbf{EmpSE} & \textbf{MSE} \\
\midrule

\textbf{No effect} \\
\quad Markov &
0.0017 & 0.0524 & 0.0027 &
0.0007 & 0.0564 & 0.0032 &
-0.0001 & 0.0527 & 0.0028 &
-0.0231 & 0.0734 & 0.0059 \\
\quad Semi-Markov &
0.0372 & 0.0527 & 0.0042 &
-0.0125 & 0.0454 & 0.0022 &
0.0005 & 0.0366 & 0.0013 &
-0.0937 & 0.0573 & 0.0121 \\
\quad State arrival extended semi-Markov &
0.0504 & 0.0583 & 0.0059 &
-0.0572 & 0.0480 & 0.0056 &
0.0276 & 0.0481 & 0.0031 &
-0.0676 & 0.0643 & 0.0087 \\
\addlinespace[3pt]

\textbf{Effect on death} \\
\quad Markov &
0.0017 & 0.0573 & 0.0033 &
0.0221 & 0.0636 & 0.0045 &
0.0004 & 0.0573 & 0.0033 &
-0.0726 & 0.0831 & 0.0122 \\
\quad Semi-Markov &
0.0431 & 0.0624 & 0.0057 &
0.0411 & 0.0541 & 0.0046 &
0.0112 & 0.0419 & 0.0019 &
-0.1240 & 0.0681 & 0.0200 \\
\quad State arrival extended semi-Markov &
0.0587 & 0.0663 & 0.0078 &
-0.0069 & 0.0566 & 0.0032 &
0.0366 & 0.0578 & 0.0047 &
-0.1030 & 0.0730 & 0.0159 \\
\addlinespace[3pt]

\textbf{Effect on disease progression \& death} \\
\quad Markov &
0.0008 & 0.0583 & 0.0034 &
0.0176 & 0.0659 & 0.0046 &
-0.0007 & 0.0582 & 0.0034 &
-0.0766 & 0.0860 & 0.0133 \\
\quad Semi-Markov &
0.0385 & 0.0684 & 0.0062 &
0.0422 & 0.0582 & 0.0052 &
0.0069 & 0.0445 & 0.0020 &
-0.1085 & 0.0746 & 0.0173 \\
\quad State arrival extended semi-Markov &
0.0493 & 0.0706 & 0.0074 &
-0.0091 & 0.0597 & 0.0036 &
0.0228 & 0.0558 & 0.0036 &
-0.0947 & 0.0790 & 0.0152 \\

\bottomrule
\end{tabular}%
}

\medskip
\noindent (B) Monte Carlo standard error (MCSE) of bias, EmpSE, and MSE

\resizebox{\textwidth}{!}{%
\begin{tabular}{l
                *{4}{>{\centering\arraybackslash}m{1.25cm}
                      >{\centering\arraybackslash}m{1.25cm}
                      >{\centering\arraybackslash}m{1.25cm}}}
\toprule
& \multicolumn{3}{c}{\textbf{Time-dependent Cox}}
& \multicolumn{3}{c}{\textbf{Landmark}}
& \multicolumn{3}{c}{\textbf{Multi-state model}}
& \multicolumn{3}{c}{\textbf{G-formula}} \\
\cmidrule(lr){2-4}\cmidrule(lr){5-7}\cmidrule(lr){8-10}\cmidrule(lr){11-13}
& \textbf{Bias} & \textbf{EmpSE} & \textbf{MSE}
& \textbf{Bias} & \textbf{EmpSE} & \textbf{MSE}
& \textbf{Bias} & \textbf{EmpSE} & \textbf{MSE}
& \textbf{Bias} & \textbf{EmpSE} & \textbf{MSE} \\
\midrule

\textbf{No effect} \\
\quad Markov &
0.0017 & 0.0012 & 0.0001 &
0.0018 & 0.0013 & 0.0001 &
0.0017 & 0.0012 & 0.0001 &
0.0023 & 0.0016 & 0.0003 \\
\quad Semi-Markov &
0.0017 & 0.0012 & 0.0002 &
0.0014 & 0.0010 & 0.0001 &
0.0012 & 0.0008 & 0.0001 &
0.0018 & 0.0013 & 0.0004 \\
\quad State arrival extended semi-Markov &
0.0018 & 0.0013 & 0.0002 &
0.0015 & 0.0011 & 0.0002 &
0.0015 & 0.0011 & 0.0001 &
0.0020 & 0.0014 & 0.0003 \\
\addlinespace[3pt]

\textbf{Effect on death} \\
\quad Markov &
0.0018 & 0.0013 & 0.0001 &
0.0020 & 0.0014 & 0.0002 &
0.0018 & 0.0013 & 0.0001 &
0.0026 & 0.0019 & 0.0005 \\
\quad Semi-Markov &
0.0020 & 0.0014 & 0.0002 &
0.0017 & 0.0012 & 0.0002 &
0.0013 & 0.0009 & 0.0001 &
0.0022 & 0.0015 & 0.0006 \\
\quad State arrival extended semi-Markov &
0.0021 & 0.0015 & 0.0003 &
0.0018 & 0.0013 & 0.0001 &
0.0018 & 0.0013 & 0.0002 &
0.0023 & 0.0016 & 0.0005 \\
\addlinespace[3pt]

\textbf{Effect on disease progression \& death} \\
\quad Markov &
0.0018 & 0.0013 & 0.0002 &
0.0021 & 0.0015 & 0.0002 &
0.0018 & 0.0013 & 0.0001 &
0.0027 & 0.0019 & 0.0005 \\
\quad Semi-Markov &
0.0022 & 0.0015 & 0.0003 &
0.0018 & 0.0013 & 0.0002 &
0.0014 & 0.0010 & 0.0001 &
0.0024 & 0.0017 & 0.0006 \\
\quad State arrival extended semi-Markov &
0.0022 & 0.0016 & 0.0003 &
0.0019 & 0.0013 & 0.0002 &
0.0018 & 0.0012 & 0.0002 &
0.0025 & 0.0018 & 0.0005 \\

\bottomrule
\end{tabular}%
}

\vspace{6pt}
\justifying
\footnotesize
\noindent Notes: $\theta$ is the true value, $\hat\theta_i$ is the estimate from replicate $i$ $(i=1,\ldots,n_{\mathrm{sim}})$, and $\bar\theta$ is their mean.
Bias $=\frac{1}{n_{\mathrm{sim}}}\sum(\hat\theta_i-\theta)$; EmpSE $=\sqrt{\frac{1}{n_{\mathrm{sim}}-1}\sum(\hat\theta_i-\bar\theta)^2}$; MSE $=\frac{1}{n_{\mathrm{sim}}}\sum(\hat\theta_i-\theta)^2$.
The Monte Carlo standard errors of Bias, EmpSE, and MSE reflect only uncertainty due to the finite number of simulation replicates; Monte Carlo variability arising from any within-estimator stochastic approximation is not separately quantified \citep{morris2019using}.
\end{table}

We then apply the four estimators to obtain the empirical RR and show the percent bias relative to the theoretical RR, computed as \(\{(\widehat{\mathrm{RR}}-\mathrm{RR}_{\mathrm{true}})/\mathrm{RR}_{\mathrm{true}}\}\times 100\%\), across the 1{,}000 replicates as box plots (each color denotes one estimator). If the estimates are unbiased, the box plots are centered at zero. Negative percent bias indicates a smaller empirical RR than the theoretical RR, that is, bias in favor of allo-HSCT, and vice versa. In the Markov setting, the time-dependent Cox model and the multi-state model are approximately unbiased, whereas the landmark supermodel shows a small bias against allo-HSCT when allo-HSCT is beneficial, and the g-formula tends to be biased in favor of allo-HSCT, in particular for earlier $\tau_{\mathrm{int}}$ in the two settings with a treatment effect. In the semi-Markov and state arrival extended semi-Markov settings, the multi-state model is generally the least biased, whereas the landmark supermodel and time-dependent Cox model may exhibit bias in different directions. The g-formula gives estimates with the largest bias, consistently favoring allo-HSCT. 
\autoref{table_2} reports the performance measures of the four estimators and the corresponding Monte Carlo standard errors for the RR of allo-HSCT scheduled at $\tau_{\mathrm{int}}=3$ months versus chemotherapy only.

We also explored alternative (non-linear) specifications of the arrival-time terms in the time-dependent Cox model to accommodate multiple time scales (in the semi-Markov and state arrival extended semi-Markov settings), as well as different choices of the parameters $\Delta s$ and \(w\) in the landmark supermodel (\refrange{Figures}{fig_3}{fig_4}). These variations did not, however, materially change the magnitude of the observed bias.

\section{Pooling deaths on a common time scale}
\label{sec.4}
The simulation results in \autoref{sec.3} show that the estimators exhibit different
degrees of bias under different process assumptions. Except for the multi-state model,
the considered approaches model the primary outcome on a single time scale and attempt
to represent additional time scales through the entry time into the current state as a
time-dependent covariate. Suppose the true process is \textit{indeed} multi-state with
distinct death transitions and, as in our deliberately simplified simulation,
three time scales are relevant (\textit{i.e.}\ time since study entry, time since disease
progression, and time since allo-HSCT). Each estimator combines a model for the
intermediate-event histories with a model for death given these histories, either of which may be misspecified
relative to the true multi-state process. For the time-dependent Cox model and the
landmark supermodel, the histories are simulated from the multi-state model
(\autoref{sec.2.4}), whose transitions and time
scales are specified correctly \textit{a priori}; their bias therefore arises from the
death model alone, which we
examine first. The g-formula models both parts itself.

We use the multi-state process $\{X(t): t\ge 0\}$ specified by $Q(s,t)$ in \autoref{sec.2.1}, and recall that $X(t)=(T_t,I_t,D_t)\in\mathcal S$, with states indexed as
\{$1\!:\!000,\, 2\!:\!010,\, 3\!:\!001,\, 4\!:\!011,\, 5\!:\!100,\, 6\!:\!110,\, 7\!:\!101,\, 8\!:\!111$\}
and allowed transitions
$\mathcal A=\{1\tto 2, 1\tto 3, 1\tto 5,\,2\tto 4,\,5\tto 6,\,5\tto 7,\,6\tto 8\}$. Consider a cohort of $n$ patients with processes $X_i(t)=(T_i(t),I_i(t),D_i(t))$, $i=1,\dots,n$. For each $(a\tto  b)\in\mathcal A$, we write $N_{i,ab}(t)$ for the transition counting process of patient $i$ and $Y_{i,a}(t)=\mathbb{I}\{X_i(t-)=a\}$ for the indicator that patient $i$ is at risk for transitions out of state $a$ at time $t$. We assume that the intensity of $N_{i,ab}(t)$ takes the transition-specific multiplicative form
\begin{equation}
Y_{i,a}(t)\,\lambda_{ab,0}(s_i, t)\exp\{\gamma_{ab}^\top Z_i\}, \qquad (a\tto b)\in\mathcal{A}.
\label{eq:2}
\end{equation}
Here, $Z_i$ denotes the baseline covariates, and $s_i=s_i(t)$ the time at which patient $i$ entered the current state $a$, with $s_i=0$ in state 1; the event history is thus captured by $a$ and $s_i$. With $h_{ab}(\cdot)$ as in \autoref{sec.2.1}, $\lambda_{ab,0}(s,t)$ equals $h_{ab}(t)$ and $h_{ab}(t-s)$ under the time-inhomogeneous Markov and semi-Markov specifications,
respectively; the state arrival extended semi-Markov specification further replaces the hazard ratio for death after progression ($2\tto 4$, $6\tto 8$) by $\rho(s)$. The decomposition below holds under all three specifications. Death can occur from the living states $\{1,2,5,6\}$ through transitions $1\tto 3$, $2\tto 4$, $5\tto 7$, and $6\tto 8$, and thus the aggregated counting process $N_{i,{\mathrm{D}}}(t)$ and the corresponding risk indicator $Y_i(t)$ are defined as
\begin{equation}
\begin{split}
N_{i,{\mathrm{D}}}(t) &:=\sum_{(a\tto b)\in\mathcal A_{\mathrm{D}}} N_{i,ab}(t), \quad \mathcal A_{\mathrm{D}}=\{1\tto 3,\,2\tto 4,\,5\tto 7,\,6\tto 8\},\\
Y_i(t) &:= \mathbb{I}\{D_i(t-)=0\}
       =\sum_{a\in \mathcal L} Y_{i,a}(t), \quad \mathcal L=\{1,2,5,6\}.
\end{split}
\label{eq:3}
\end{equation}

A time-dependent Cox model treats death as a single event process and does
not distinguish the originating states. In the counting-process notation above, the
hazard model of \autoref{sec.2.3} specifies the intensity of $N_{i,\mathrm{D}}(t)$
as 
\begin{equation}
Y_i(t)\,h_0(t)\exp\{\beta^\top Z_i(t)\},
\label{eq:cox}
\end{equation}
with $Z_i(t)=\big(T_i(t-),\,I_i(t-),\,Z_i\big)$ and
   $\beta=(\beta_\mathrm{T},\beta_\mathrm{I},\beta_\mathrm{Z})$.
The partial likelihood score in the Cox model is therefore
\begin{equation}
U(\beta)
=
\sum_{i=1}^n\int_{0}^{t^*}\Bigl(Z_i(t)-\bar Z(\beta,t)\Bigr)\,dN_{i,\mathrm{D}}(t)
=
\sum_{(a\tto b)\in\mathcal A_{\mathrm{D}}}\sum_{i=1}^n\int_{0}^{t^*}\Bigl(Z_i(t)-\bar Z(\beta,t)\Bigr)\,dN_{i,ab}(t),
\label{eq:4}
\end{equation}
where $\bar Z(\beta,t)$ is the exp-weighted mean of $Z_j(t)$ over the pooled risk set
at time $t$,
\begin{equation}
\bar Z(\beta,t)=\frac{\sum_{j=1}^n Y_j(t)\exp\{\beta^\top Z_j(t)\}\,Z_j(t)}
                     {\sum_{j=1}^n Y_j(t)\exp\{\beta^\top Z_j(t)\}}.
\label{eq:5}
\end{equation}
By \eqref{eq:4}, deaths from different originating states contribute additively to the score and
are all centered by the same $\bar Z(\beta,t)$. We next decompose
the score by originating state. For each $a\in\mathcal L$, we define
the exp-weighted mean of $Z_j(t)$ over the risk set of state $a$ and the proportion of the exp-weighted pooled risk set in state $a$ as
\begin{equation}
\bar Z_a(\beta,t)
=
\frac{\sum_{j=1}^n Y_{j,a}(t)\exp\{\beta^\top Z_j(t)\}\,Z_j(t)}
     {\sum_{j=1}^n Y_{j,a}(t)\exp\{\beta^\top Z_j(t)\}},
\qquad
w_a(\beta,t)
=
\frac{\sum_{j=1}^n Y_{j,a}(t)\exp\{\beta^\top Z_j(t)\}}
     {\sum_{j=1}^n Y_j(t)\exp\{\beta^\top Z_j(t)\}}.
\label{eq:6}
\end{equation}
Since $Y_j(t)=\sum_{a\in\mathcal L}Y_{j,a}(t)$ by \eqref{eq:3}, we have
\begin{equation}
\bar Z(\beta,t)=\sum_{a\in\mathcal L} w_a(\beta,t)\,\bar Z_a(\beta,t).
\label{eq:7}
\end{equation}
Writing $Z_i(t)-\bar Z(\beta,t)=\{Z_i(t)-\bar Z_a(\beta,t)\}+\{\bar Z_a(\beta,t)-\bar Z(\beta,t)\}$
for deaths from state $a$, the score \eqref{eq:4} becomes
\begin{align}
U(\beta)
&=
\sum_{(a\tto b)\in\mathcal A_{\mathrm{D}}}\sum_{i=1}^n\int_0^{t^*}\Big(Z_i(t)-\bar Z_a(\beta,t)\Big)\,dN_{i,ab}(t)
\nonumber\\
&\quad+
\sum_{(a\tto b)\in\mathcal A_{\mathrm{D}}}\int_0^{t^*}\Big(\bar Z_a(\beta,t)-\bar Z(\beta,t)\Big)\,dN_{\cdot,ab}(t),
\quad
dN_{\cdot,ab}(t)=\sum_{i=1}^n dN_{i,ab}(t).
\label{eq:8}
\end{align}
The first term is the score of a Cox model stratified by originating state, that is, of a
multi-state model with transition-specific baselines and common covariate effects; the
second term arises only from pooling and compares the state-specific means
$\bar Z_a(\beta,t)$ with the pooled mean $\bar Z(\beta,t)$, weighted by the observed
deaths from each state. Since all individuals in a given state share the same values of $T_i(t-)$ and $I_i(t-)$, the corresponding components of $Z_i(t)-\bar Z_a(\beta,t)$ in the first term of \eqref{eq:8} are zero, so that $\beta_\mathrm{T}$ and $\beta_\mathrm{I}$ are determined
entirely by the second term, that is, by contrasts between the death rates of different
originating states at the same time $t$. 

To illustrate, consider the simplest scenario without disease progression and ignore
covariates, in which case deaths occur only through $1\tto3$ and $5\tto7$, the living states
are $a\in\{1,5\}$, and $Z_i(t)=T_i(t-)$. Then $\bar Z_1(\beta,t)=0$,
$\bar Z_5(\beta,t)=1$, and $\bar Z(\beta,t)=w_5(\beta,t)$ by \eqref{eq:7}, so that the
second term of \eqref{eq:8} reduces to
\begin{equation*}
\int_0^{t^*}\Big\{\big(1-w_5(\beta,t)\big)\,dN_{\cdot,57}(t)-w_5(\beta,t)\,dN_{\cdot,13}(t)\Big\}.
\end{equation*}
By \eqref{eq:6}, $w_5(\beta,t)=Y_{\cdot,5}(t)e^{\beta_\mathrm{T}}/\{Y_{\cdot,1}(t)+Y_{\cdot,5}(t)e^{\beta_\mathrm{T}}\}$, where $Y_{\cdot,1}(t)$ and $Y_{\cdot,5}(t)$ denote the numbers at risk in
states 1 and 5, and the expectations of $dN_{\cdot,57}(t)$ and $dN_{\cdot,13}(t)$ are
$Y_{\cdot,5}(t)\lambda_{57}(t)\,dt$ and $Y_{\cdot,1}(t)\lambda_{13}(t)\,dt$. The expected integrand at time $t$ is therefore
\begin{equation*}
\frac{Y_{\cdot,1}(t)\,Y_{\cdot,5}(t)}{Y_{\cdot,1}(t)+Y_{\cdot,5}(t)e^{\beta_\mathrm{T}}}
\bigl\{\lambda_{57}(t)-e^{\beta_\mathrm{T}}\lambda_{13}(t)\bigr\}\,dt,
\end{equation*}
which vanishes only if $e^{\beta_\mathrm{T}}$ equals the ratio $\lambda_{57}(t)/\lambda_{13}(t)$
of post- to pre-transplant death rates at that time. If this ratio is constant in $t$, a single $\beta_\mathrm{T}$ sets the integrand to zero at
every $t$; otherwise, no single $\beta_\mathrm{T}$ does so, and $e^{\hat\beta_\mathrm{T}}$
effectively averages the ratio over time.

Beyond this simple case, whether these contrasts can be captured by the constant
coefficients of the time-dependent Cox model depends on whether the death intensities satisfy
\begin{equation}
\lambda_{ab,0}(s,t)\exp\{\gamma_{ab}^\top Z_i\}
= \lambda_{\mathrm{D},0}(t)\exp\{\beta_\mathrm{T} T_a + \beta_\mathrm{I} I_a + \beta_\mathrm{Z}^\top Z_i + g(s)\},
\label{eq:9}
\end{equation}
for some common baseline $\lambda_{\mathrm{D},0}(t)$ and function $g(\cdot)$, where
$(T_a,I_a)$ are the allo-HSCT and progression components of state $a$. In other words, the death intensities from different
originating states must be proportional on the common time scale $t$, up to a factor
depending on the arrival time only; in the example above, this reduces to a constant
ratio $\lambda_{57}(t)/\lambda_{13}(t)$. 
Under \eqref{eq:9}, summing the death intensities \eqref{eq:2} over
$(a\tto b)\in\mathcal A_\mathrm{D}$ gives
\[
Y_i(t)\,\lambda_{\mathrm{D},0}(t)\exp\{\beta^\top Z_i(t)+g(s_i)\},
\]
which is exactly the
intensity specified by the time-dependent Cox model \eqref{eq:cox}, augmented by the entry-time term
$g(u)$ where needed, with $h_0(t)=\lambda_{\mathrm{D},0}(t)$. The model is then correctly
specified, and both terms in \eqref{eq:8} have mean zero at the true parameters. If \eqref{eq:9} fails, as in the example when the ratio varies with $t$,
$\hat\beta$ converges to a pseudo-true `compromise' value, which depends on how the pooled risk set
mixes individuals across living states over time through $w_a(\beta,t)$, the exp-weighted share of state $a$ in the pooled risk set.

We now relate \eqref{eq:9} to the specifications in \autoref{table_1}. All death intensities under the time-inhomogeneous Markov specification are multiples of $h_{13}(t)$, so that \eqref{eq:9} holds with $g(\cdot)\equiv 0$. The time-dependent Cox model is then correctly specified and, with the intermediate-event histories simulated from the multi-state model, yields approximately unbiased RRs (\autoref{fig_2}). Under clock reset with Weibull hazards, as in our simulation, however, the post-event death intensities become multiples of $h_{13}(t-s)$ on the common time scale $t$, and $h_{13}(t-s)/h_{13}(t)$ is not a function of $s$ alone; \eqref{eq:9} thus fails for any $g(\cdot)$, with or without the factor $\rho(s)$. The time-dependent Cox model then converges to a pseudo-true compromise value, consistent with its bias in \autoref{fig_2}, which is insensitive to the specification of $g(u)$ (\autoref{fig_3}). We note that whether \eqref{eq:9} holds depends not on the clock reset as such, but on whether the post-event intensities can still be written on the time scale $t$ with a multiplicative factor depending only on the arrival time. For exponential hazards, clock reset is immaterial, and, for Gompertz hazards $h(t)=a\,e^{bt}$, we have $h(t-s)=h(t)\,e^{-bs}$, so that clock reset amounts to a state arrival extended Markov specification with a log-linear arrival-time factor, and \eqref{eq:9} holds with linear $g(\cdot)$. Such a specification is natural when the subsequent hazard retains its dependence on time since study entry but is modified multiplicatively by the timing (or extent) of an intermediate event, for example the duration of atrial fibrillation for subsequent stroke risk \citep{kaplan2019stroke}.

Recall from \autoref{sec.2.3} that a landmark supermodel stacks the landmark datasets,
each comprising the risk set $\mathcal R(s)=\{i:Y_i(s)=1\}$ with follow-up in $(s,s+w]$
and covariates $Z_i(s)$ fixed at the landmark time $s$, and fits a single Cox model with
landmark-dependent effects $\beta(s)$ and a baseline hazard that varies with $s$ through
$\exp\{g(s)\}$. Two features distinguish it from the time-dependent Cox model: the state
indicators are frozen at $s$, and each death is centered by an exp-weighted mean over all
rows at risk at time $t$, which pools individuals across both landmark datasets and
originating states. Under a correctly specified time-dependent Cox model and a Markov process,
\citet{putter2017understanding} showed for a single landmark that freezing the treatment
indicator at $s$ attenuates its coefficient towards zero: patients transplanted within
$(s,s+w]$ remain coded as untransplanted, a misclassification of person-time that mirrors
that of an ever/never treatment indicator in a time-fixed Cox model, but in the opposite direction. When allo-HSCT is
beneficial, the landmark estimates are thus biased against allo-HSCT, consistent with
\autoref{fig_2} under the Markov specification. Our setting goes beyond theirs in two respects. First, the stacked supermodel combines the
landmark-specific contrasts across $s$ through the parameterization of $\beta(s)$ and
$g(s)$. Second, under clock reset, the misclassification from freezing persists and, as for the time-dependent Cox model, \eqref{eq:9} fails, so that the bias of the landmark supermodel reflects both mechanisms. Notably, the landmark supermodel was originally proposed to target conditional, prediction-oriented contrasts given survival
and state history at $s$, which need not coincide with causal estimands defined at baseline.

Like the multi-state model, the g-formula models the intermediate-event histories
itself. When the within-interval death probabilities
$p_{\mathrm D,k}=\Pr(D_k=1\mid\mathcal H_{k-1})$ are small,
$\operatorname{logit}p_{\mathrm{D},k}\approx\log p_{\mathrm{D},k}\approx\log\Lambda_{\mathrm{D},k}+\eta(\mathcal H_{k-1})$
for a multiplicative continuous-time hazard with linear predictor $\eta(\mathcal H_{k-1})$
and cumulative baseline $\Lambda_{\mathrm{D},k}$ over $[t_{k-1},t_k)$, so that the pooled
logistic death model approximates a proportional hazards model on the time scale since study entry
\citep{d1990relation}. In our simulation, the models for disease progression and death, specified identically
under all three processes (\autoref{fig_2}), summarize the allo-HSCT and progression
histories and include a linear term in
$t_{k-1}$; their log baselines are thus linear in $t_{k-1}$. This, unlike the unspecified $h_0(t)$ in the Cox model, cannot reproduce the
Weibull baselines in \autoref{table_1}, whose logarithms are linear in $\log t$; the resulting misspecification contributes to the bias of the g-formula even under the Markov
specification; a more flexible time trend may reduce it, although we did not examine this here. The log-logistic death
hazards in \hyperref[app.b]{Appendix~B} are likewise not log-linear in $t$, but less
markedly so, consistent with the smaller bias of the g-formula under the Markov
setting (\autoref{fig_b1}). Fitted to person-intervals pooled across originating states, the death model is moreover
subject to \eqref{eq:9}. The progression model, which generates the simulated histories,
is misspecified for the same reasons. This dual source of misspecification is consistent
with the largest bias observed for the g-formula (\autoref{fig_2}).

\section{Treatment beyond a time-dependent covariate}
\label{sec.5}
In \autoref{sec.4}, we discussed how pooling deaths from different originating states on a common time scale results in pseudo-true compromise model parameters: these reflect associational, rather than causal, `current-state' effects, which are not entirely attributable to the treatment (\textit{i.e.}\ allo-HSCT) itself, but are essentially pooled across these states and depend on modeling constructs, such as the time of entering a new state or landmark time points. Beyond the resulting bias, this raises
the question of what such coefficients represent and whether they can inform treatment
decisions at all.

In this regard, we refer to the work of \cite{van2008dynamic, van2011dynamic} on the landmark supermodel (with similar considerations applying to the time-dependent Cox model and the g-formula). The authors proposed the landmark supermodel as a potential alternative to a multi-state model for dynamic prediction. They noted that the landmark supermodel offers the advantage of model parsimony (\textit{i.e.}\ direct modeling using simple regression models rather than transition-specific models for the outcome), although it does not provide biological insight or causal interpretation. It is worth noting that the original application of the landmarking approach was to predict \textit{post-transplant} survival for patients with acute lymphoid leukemia, assessing prognostic factors such as acute graft-versus-host disease and platelet recovery. Here, the dynamic changes in these intermediate disease markers can be viewed as surrogates reflecting the underlying disease dynamics, and associational effects that do not separate the markers from the accompanying state changes suffice for the predictive objective of understanding future disease prognosis. Such a perspective, however, differs from causal effect estimation, which compares the outcomes between the counterfactual worlds in which a patient would, versus would not, undergo the treatment of interest. In this setting, a model parameter is used to summarize the effect of a treatment that is
not a purely internal disease marker evolving along the natural course, but an external
intervention imposed on the disease process. Attributing effects arising from state changes to the treatment of interest can therefore lead to misleading conclusions and, ultimately, suboptimal therapeutic decisions. Importantly, in the context of immortal time bias, the target is typically a causal effect estimand rather than a predictive contrast under the observed treatment practice. Such a target can also be viewed as a prediction under a hypothetical intervention
\citep{van2020prediction}, that is, a risk under a hypothetical treatment strategy,
as required for treatment decisions at baseline.

That said, the treatment of interest is also often not purely external to the disease process. For example, in our case, allo-HSCT eligibility is subject to disease progression over time. This implies that merely comparing risks by toggling the value of the binary,
time-dependent treatment indicator in an outcome model targets a contrast that is
generally not identifiable, as positivity is structurally violated, and is prone to
collider bias through conditioning on post-baseline eligibility and survival. As noted in \autoref{sec.1}, this relates to the long-standing question of whether the hazard ratio allows for a causal interpretation \citep{hernan2010hazards, martinussen2022causality, beyersmann2025hazards}. We also draw on the \cite{gerstung2017precision} knowledge bank approach, a multi-state model--based therapeutic decision support tool for assessing allo-HSCT versus chemotherapy treatment effects, which substantially inspired our work. However, it contrasts predicted risks under different values of the time-dependent treatment indicator, which does not, by itself, resolve the identification issues that arise when eligibility evolves endogenously over time. 

In our target trial emulation (\autoref{sec.2.4}), the intermediate-event histories are instead simulated under each strategy from the multi-state model, so that allo-HSCT occurs at $\tau_\mathrm{int}$ only in patients who are alive and progression-free at that time, and the treatment indicator follows the simulated disease process rather than being set irrespective of eligibility. This brings the time-dependent treatment back to the setting considered by \cite{beyersmann2025hazards}, in which causal effects are contrasts between interventional distributions indexed by a treatment intention at the time origin: each strategy is such an intention, and the RR contrasts the resulting interventional
distributions of the multi-state process, with the identification conditions of \autoref{sec.2.2} taking the place of randomization. The coefficients discussed in \autoref{sec.4} can thus inform treatment decisions only through the strategy-specific risks they imply, and only if the underlying models reflect the disease process.

\section{Illustration: AMLSG trial data}
\label{sec.aml}

To illustrate how these estimators compare in a realistic setting, we used the de-identified data of 1,540 adults with AML from three prospective AMLSG trials (AML-HD 98A, AML-HD 98B, and AMLSG 07-04; \citealp{papaemmanuil2016genomic}), made publicly available by \cite{gerstung2017precision}. In these trials, allo-HSCT in first CR was based on genetic risk and donor availability. Study entry was defined as the date of first CR, at which early response is established (\autoref{sec.2.1}). Of the 1,269 patients achieving CR, we excluded 120 who had been transplanted before CR was documented, mostly for primary refractory disease%
, and 86 with missing ELN risk, leaving 1,063 patients. As in the simulation, we restricted the baseline covariates to two prognostic factors, age (continuous; median 48 years, range 18--80) and ELN risk (428 favorable, 501 intermediate, and 134 adverse), and truncated follow-up at 36 months. Within this period, 323 patients received allo-HSCT in first CR, 490 experienced disease progression (\textit{i.e.}\ relapse from first CR), and 344 died. Allo-HSCT took place at a median of 2.6 months after first CR (interquartile range 1.5--4.1), and within 6 months in 308 of the 323 patients; we therefore considered strategies scheduling allo-HSCT at $\tau_{\mathrm{int}} \in \{1, 2, 3, 6\}$ months. We defined both strategies to exclude allo-HSCT after relapse ($\lambda_{26} = 0$); the 208 such transplants (including 83 performed in a second or later CR) were therefore treated as censoring events, under the assumption that censoring is independent of the mortality hazard conditional on age and ELN risk. Since salvage transplantation is likely to select patients with a more favorable post-relapse prognosis, this assumption may overestimate post-relapse mortality and hence the risk, particularly under the chemotherapy-only strategy. As the same censoring assumption applies to all four estimators, it should not affect their comparison, which is the purpose of this illustration. Of the remaining identification conditions outlined in \autoref{sec.2.2}, consistency requires both strategies to be well defined; we define them as implemented in the trials, that is, allo-HSCT in first CR, irrespective of donor source and conditioning regimen, versus protocol consolidation, including autologous HSCT in 88 patients; positivity is likely to be supported for the strategies considered, as almost all transplants took place within 6 months and occurred in all ELN risk groups and in both younger and older patients; conditional exchangeability given age and ELN risk alone is, however, unlikely to hold, since transplant decisions were based on a comprehensive clinical assessment and thus also likely depended on other prognostic factors not adjusted for here. The results presented in this section should therefore not be interpreted as clinical effect estimates.

The estimators were specified as in \autoref{sec.3}, except that the time-dependent Cox
model used quadratic instead of linear arrival-time terms; under the state arrival extended semi-Markov specification, the multi-state model additionally included the time of relapse as a linear covariate for death after relapse ($2 \tto 4$, $6 \tto 8$), since spline terms with varying degrees of freedom did not improve the model fit. Counterfactual intermediate-event histories were simulated from the fitted multi-state model with 50 trajectories per patient. Uncertainty was quantified by 95\% percentile bootstrap confidence intervals (CIs) based on 500 resamples of patients, with the entire procedure repeated in each resample (see \autoref{tab:aml} for details).

\begin{table}[htbp]
\centering
\caption{Estimated absolute risk of death by 36 months under the chemotherapy-only and intention-to-transplant strategies in the AMLSG data ($n = 1{,}063$): point estimates with 95\% percentile bootstrap confidence intervals}
\label{tab:aml}
\small
\begin{tabular}{@{}p{0.22\linewidth}ccc@{}}
\toprule
 & \multicolumn{3}{c}{Time-scale specification of the multi-state process} \\
\cmidrule(l){2-4}
Estimator & \makebox[\dimexpr0.26\linewidth-2\tabcolsep\relax]{Markov} & \makebox[\dimexpr0.26\linewidth-2\tabcolsep\relax]{Semi-Markov} & \makebox[\dimexpr0.26\linewidth-2\tabcolsep\relax]{\shortstack{State arrival\\ extended semi-Markov}} \\
\midrule
\multicolumn{4}{@{}l}{\textit{Chemotherapy only (no allo-HSCT)}} \\
\quad Time-dependent Cox & 0.464 (0.419, 0.504) & 0.460 (0.415, 0.501) & 0.460 (0.415, 0.501) \\
\quad Landmark & 0.485 (0.432, 0.530) & 0.483 (0.432, 0.527) & 0.483 (0.432, 0.527) \\
\quad Multi-state model & 0.462 (0.411, 0.500) & 0.463 (0.414, 0.504) & 0.455 (0.408, 0.495) \\
\quad G-formula$^{a}$ & 0.451 (0.415, 0.487) & 0.451 (0.415, 0.487) & 0.451 (0.415, 0.487) \\
\addlinespace
\multicolumn{4}{@{}l}{\textit{Allo-HSCT scheduled at $\tau_{\mathrm{int}} = 1$ month}} \\
\quad Time-dependent Cox & 0.347 (0.305, 0.398) & 0.301 (0.242, 0.362) & 0.301 (0.242, 0.362) \\
\quad Landmark & 0.351 (0.305, 0.402) & 0.351 (0.303, 0.402) & 0.351 (0.303, 0.402) \\
\quad Multi-state model & 0.346 (0.296, 0.406) & 0.327 (0.285, 0.385) & 0.324 (0.280, 0.382) \\
\quad G-formula$^{a}$ & 0.339 (0.284, 0.396) & 0.339 (0.284, 0.396) & 0.339 (0.284, 0.396) \\
\addlinespace
\multicolumn{4}{@{}l}{\textit{Allo-HSCT scheduled at $\tau_{\mathrm{int}} = 2$ months}} \\
\quad Time-dependent Cox & 0.350 (0.309, 0.396) & 0.326 (0.283, 0.379) & 0.326 (0.283, 0.379) \\
\quad Landmark & 0.359 (0.314, 0.407) & 0.358 (0.312, 0.406) & 0.358 (0.312, 0.406) \\
\quad Multi-state model & 0.346 (0.300, 0.404) & 0.338 (0.296, 0.393) & 0.335 (0.294, 0.391) \\
\quad G-formula$^{a}$ & 0.373 (0.332, 0.420) & 0.373 (0.332, 0.420) & 0.373 (0.332, 0.420) \\
\addlinespace
\multicolumn{4}{@{}l}{\textit{Allo-HSCT scheduled at $\tau_{\mathrm{int}} = 3$ months}} \\
\quad Time-dependent Cox & 0.354 (0.316, 0.398) & 0.359 (0.316, 0.414) & 0.359 (0.316, 0.414) \\
\quad Landmark & 0.367 (0.326, 0.413) & 0.370 (0.329, 0.418) & 0.370 (0.329, 0.418) \\
\quad Multi-state model & 0.349 (0.306, 0.402) & 0.353 (0.312, 0.408) & 0.354 (0.312, 0.408) \\
\quad G-formula$^{a}$ & 0.400 (0.361, 0.443) & 0.400 (0.361, 0.443) & 0.400 (0.361, 0.443) \\
\addlinespace
\multicolumn{4}{@{}l}{\textit{Allo-HSCT scheduled at $\tau_{\mathrm{int}} = 6$ months}} \\
\quad Time-dependent Cox & 0.383 (0.348, 0.418) & 0.467 (0.391, 0.529) & 0.467 (0.391, 0.529) \\
\quad Landmark & 0.400 (0.363, 0.438) & 0.408 (0.368, 0.448) & 0.408 (0.368, 0.448) \\
\quad Multi-state model & 0.376 (0.337, 0.417) & 0.397 (0.359, 0.445) & 0.406 (0.367, 0.454) \\
\quad G-formula$^{a}$ & 0.452 (0.412, 0.493) & 0.452 (0.412, 0.493) & 0.452 (0.412, 0.493) \\
\bottomrule
\multicolumn{4}{@{}p{\linewidth}@{}}{\footnotesize $^{a}$ The g-formula does not use the time-scale specification; the same estimate is shown in all columns. \newline Time-dependent Cox with quadratic arrival-time terms in the semi-Markov and state arrival extended semi-Markov columns; landmark model with monthly landmarks and a 6-month horizon. \newline Simulated trajectories per patient: 50 (point estimates and each bootstrap resample). Bootstrap resamples: 499 of 500 (one excluded because the multi-state fit failed); g-formula: 500 within \textit{gfoRmula}.} \\
\end{tabular}
\end{table}

\begin{table}[htbp]
\centering
\caption{Risk ratio of death by 36 months relative to the chemotherapy-only strategy (no allo-HSCT) under intention-to-transplant strategies in the AMLSG data ($n = 1{,}063$): point estimates with 95\% percentile bootstrap confidence intervals}
\label{tab:aml_rr}
\small
\begin{tabular}{@{}p{0.22\linewidth}ccc@{}}
\toprule
 & \multicolumn{3}{c}{Time-scale specification of the multi-state process} \\
\cmidrule(l){2-4}
Estimator & \makebox[\dimexpr0.26\linewidth-2\tabcolsep\relax]{Markov} & \makebox[\dimexpr0.26\linewidth-2\tabcolsep\relax]{Semi-Markov} & \makebox[\dimexpr0.26\linewidth-2\tabcolsep\relax]{\shortstack{State arrival\\ extended semi-Markov}} \\
\midrule
\multicolumn{4}{@{}l}{\textit{Allo-HSCT scheduled at $\tau_{\mathrm{int}} = 1$ month}} \\
\quad Time-dependent Cox & 0.748 (0.650, 0.874) & 0.654 (0.519, 0.815) & 0.654 (0.519, 0.815) \\
\quad Landmark & 0.724 (0.601, 0.858) & 0.726 (0.601, 0.857) & 0.726 (0.601, 0.857) \\
\quad Multi-state model & 0.748 (0.629, 0.905) & 0.707 (0.597, 0.854) & 0.712 (0.602, 0.850) \\
\quad G-formula$^{a}$ & 0.751 (0.625, 0.910) & 0.751 (0.625, 0.910) & 0.751 (0.625, 0.910) \\
\addlinespace
\multicolumn{4}{@{}l}{\textit{Allo-HSCT scheduled at $\tau_{\mathrm{int}} = 2$ months}} \\
\quad Time-dependent Cox & 0.754 (0.664, 0.873) & 0.709 (0.604, 0.844) & 0.709 (0.604, 0.844) \\
\quad Landmark & 0.739 (0.623, 0.871) & 0.741 (0.624, 0.868) & 0.741 (0.624, 0.868) \\
\quad Multi-state model & 0.749 (0.636, 0.896) & 0.729 (0.626, 0.870) & 0.738 (0.633, 0.874) \\
\quad G-formula$^{a}$ & 0.829 (0.724, 0.956) & 0.829 (0.724, 0.956) & 0.829 (0.724, 0.956) \\
\addlinespace
\multicolumn{4}{@{}l}{\textit{Allo-HSCT scheduled at $\tau_{\mathrm{int}} = 3$ months}} \\
\quad Time-dependent Cox & 0.763 (0.678, 0.879) & 0.780 (0.682, 0.925) & 0.780 (0.682, 0.925) \\
\quad Landmark & 0.757 (0.649, 0.885) & 0.767 (0.657, 0.897) & 0.767 (0.657, 0.897) \\
\quad Multi-state model & 0.755 (0.655, 0.885) & 0.763 (0.662, 0.898) & 0.779 (0.674, 0.905) \\
\quad G-formula$^{a}$ & 0.888 (0.796, 0.996) & 0.888 (0.796, 0.996) & 0.888 (0.796, 0.996) \\
\addlinespace
\multicolumn{4}{@{}l}{\textit{Allo-HSCT scheduled at $\tau_{\mathrm{int}} = 6$ months}} \\
\quad Time-dependent Cox & 0.824 (0.752, 0.911) & 1.015 (0.849, 1.181) & 1.015 (0.849, 1.181) \\
\quad Landmark & 0.825 (0.735, 0.932) & 0.844 (0.749, 0.960) & 0.844 (0.749, 0.960) \\
\quad Multi-state model & 0.813 (0.736, 0.917) & 0.857 (0.770, 0.975) & 0.894 (0.806, 1.011) \\
\quad G-formula$^{a}$ & 1.002 (0.916, 1.101) & 1.002 (0.916, 1.101) & 1.002 (0.916, 1.101) \\
\bottomrule
\multicolumn{4}{@{}p{\linewidth}@{}}{\footnotesize $^{a}$ The g-formula does not use the time-scale specification; the same estimate is shown in all columns. Specifications and bootstrap as in \autoref{tab:aml}.} \\
\end{tabular}
\end{table}

\begin{figure}[htbp]
\centering
\includegraphics[width=\textwidth]{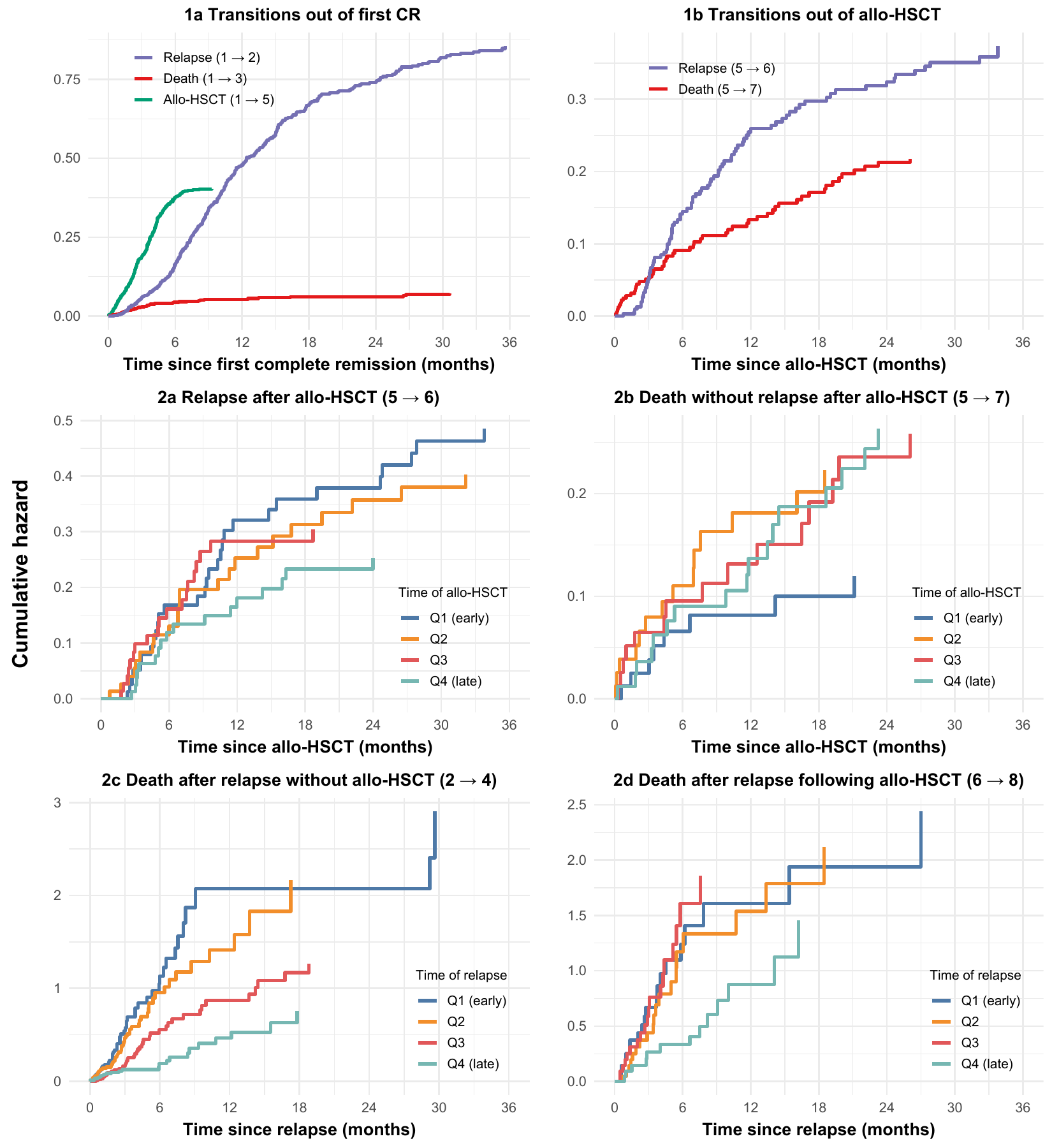}
\caption{Nelson--Aalen estimates of the cumulative transition hazards in the AMLSG data ($n = 1{,}063$), truncated at 36 months after study entry (first CR). (1a) Transitions out of first CR to relapse ($1 \tto 2$), death ($1 \tto 3$), and allo-HSCT ($1 \tto 5$), by time since first CR. (1b) Transitions out of allo-HSCT to relapse ($5 \tto 6$) and death ($5 \tto 7$), by time since allo-HSCT. (2a) Relapse after allo-HSCT ($5 \tto 6$) and (2b) death without relapse after allo-HSCT ($5 \tto 7$), by time since allo-HSCT, stratified by quartiles of the time of allo-HSCT. (2c) Death after relapse without allo-HSCT ($2 \tto 4$) and (2d) death after relapse following allo-HSCT ($6 \tto 8$), by time since relapse, stratified by quartiles of the time of relapse. Estimates are not adjusted for baseline covariates.}
\label{fig:aml_cumhaz}
\end{figure}

\autoref{fig:aml_cumhaz} shows Nelson--Aalen estimates of the cumulative transition hazards and provides a simple graphical diagnostic of the time-scale assumptions in \autoref{sec.2.1}. Panels 1a and 1b display the transitions out of states 1 and 5 against the time since first CR and since allo-HSCT, respectively. Allo-HSCT occurred almost exclusively within the first 9 months (Panel 1a). Compared with Panel 1a, the unadjusted cumulative hazard of relapse after allo-HSCT was markedly lower, consistent with an anti-leukemic effect, but had a similar shape on its own time scale, rising steeply after a short initial delay before leveling off (Panel 1b). Death without relapse, in contrast, was more common after allo-HSCT, consistent with transplant-related mortality, and its cumulative hazard increased most steeply immediately after transplantation, even though transplants took place at different times after first CR; this peri-transplant excess hazard therefore appears anchored at the time of transplantation rather than at study entry. These observations thus favor a clock reset over the time-inhomogeneous Markov specification. Panels 2a and 2b show the transitions out of state 5 against the time since allo-HSCT, stratified by quartiles of allo-HSCT time, and Panels 2c and 2d the death transitions out of states 2 and 6 against the time since relapse, stratified by quartiles of relapse time. Under a semi-Markov process, these stratified curves should coincide, whereas a systematic separation by arrival time points to a dependence on the time since study entry, as under a time-inhomogeneous Markov process, or on the arrival time itself, as under a state arrival extended semi-Markov process; the plots alone cannot distinguish between the two. The cumulative hazards of relapse and death after allo-HSCT were broadly similar across quartiles of allo-HSCT time (Panels 2a, 2b), whereas mortality after relapse, with or without prior allo-HSCT, was markedly higher after early than after late relapse (Panels 2c, 2d). This separation may reflect more aggressive disease in patients who relapse early, in line with the established prognostic value of remission duration for survival after relapse \citep{breems2005prognostic, schmid2012treatment, gerstung2017precision}. Since the curves suggest a clock reset after allo-HSCT, we regard the state arrival extended semi-Markov specification as the most plausible. As these descriptive plots provide no formal test, we report results under all three specifications.

Tables~\ref{tab:aml} and~\ref{tab:aml_rr} report the estimated 3-year mortality risks under each strategy and the corresponding RRs relative to chemotherapy only, with 95\% bootstrap CIs. The risk under chemotherapy only was similar across estimators and specifications (0.451--0.485). Under the Markov specification, the time-dependent Cox model, landmark supermodel, and multi-state model yielded nearly identical RRs (0.724--0.748 for allo-HSCT scheduled at 1 month and 0.813--0.825 at 6 months); the g-formula agreed at 1 month (0.751) but attenuated faster, reaching 1.002 at 6 months. Under the non-Markov specifications, the estimators diverged: the time-dependent Cox model depended most strongly on the scheduling time (0.654 at 1 month and 1.015 at 6 months), the landmark supermodel least (0.726 at 1 month and 0.844 at 6 months), and the multi-state model lay in between (0.707--0.712 at 1 month and 0.857--0.894 at 6 months). As the two non-Markov specifications differ only in death after relapse, they generate identical counterfactual intermediate-event histories, so that the time-dependent Cox model and landmark supermodel give identical estimates under both. At 6 months under the state arrival extended semi-Markov specification, the estimated relative reduction in 3-year mortality ranged from about 16\% (landmark supermodel) to none (time-dependent Cox model). The bootstrap CIs indicate considerable sampling uncertainty; at 6 months, whether they excluded one depended on the estimator and the assumed time scale.

All estimators used the same data and the same strategy definitions, and, under a given specification, all except the g-formula used the same counterfactual intermediate-event histories. Because the true RR is unknown in this application, these results cannot identify the least biased estimator. They do show, however, that, in a realistic setting, the choice of estimator and assumed time scale alone can change the estimated benefit of allo-HSCT quantitatively and, for later scheduling times, even qualitatively. Under the state arrival extended semi-Markov specification, which we regard as the most plausible, the simulation results of \autoref{sec.3} favor the multi-state estimates. This conclusion, however, rests on the assumed process and on the Weibull mortality hazards in the simulation, and cannot be verified from the data. With log-logistic mortality hazards (\hyperref[app.b]{Appendix~B}), the time-dependent Cox model was instead the least biased under this specification (\autoref{fig_b1}).

For applied analyses of similar questions, these results suggest three practical steps. First, the estimand should be clearly defined before the analysis, including the time origin, the treatment strategies and their handling of intercurrent events, the time horizon, and a summary measure on the probability scale. Second, estimating strategy-specific risks often requires emulating a target trial: loss of eligibility before the scheduled treatment is part of the strategy rather than a reason for exclusion, and only strategies supported by the data should be considered, as estimates for poorly supported strategies (here, for instance, scheduling beyond 6 months) rely on extrapolation beyond the data. Third, assumptions about the underlying process should be examined with diagnostic tools such as the plots in \autoref{fig:aml_cumhaz}, together with clinical knowledge of the disease process, rather than inferred from agreement between estimators. Our simulations imply that, if the disease process is as assumed, the multi-state model is generally the least biased under Weibull hazards; its estimates are therefore only as reliable as the assumed process. In practice, several estimators are often reported together as a sensitivity analysis, with their agreement taken as support for the conclusions, but, as shown above, such agreement may reflect shared assumptions rather than correct estimates: under the Markov specification, the time-dependent Cox model, landmark supermodel, and multi-state model yielded nearly identical RRs, but they diverged once a clock reset was assumed.

\section{Discussion}
\label{sec.6}

Many papers report treatment effects of transplantation as hazard ratios \citep{nilsson2019secondary,devillier2022depth}, which are often poorly aligned with therapeutic decisions at baseline in settings prone to immortal time bias. Translating such estimates to the probability scale for clinical decision making, as attempted by \cite{gerstung2017precision} in a multi-state setting, remains rare in practice. We discuss this problem in the context of AML, but the implications are much broader: the problem may arise whenever randomized controlled trials are not feasible and interventions substantially alter patients' risk profiles (\textit{e.g.}\ solid organ transplantation or high-risk cardiac procedures), thereby introducing additional time scales.

This leads to the next point: when the aim is an unbiased causal effect estimate to inform the decision for a patient who may be eligible for a specific therapy later, it is important to reconsider whether multiple time scales are required and, in particular, whether a Markov assumption is defensible. Collapsing the process onto a common time scale is computationally simple but risks misspecification that cannot be easily remedied by more flexible functional forms or covariate specifications. We regard multi-state models as key to describing such dynamics; robust diagnostic tools to assess their adequacy and to identify dynamic patterns in real-world data are an important direction for future work \citep{drenda2026comparison}.

We designed a phase III simulation study to mimic AML patient trajectories in observational settings. Based on these trajectories, we adopted causal approaches to assess the performance of different estimators. We demonstrated that neither the CTD models nor the g-formula can, by default, guarantee unbiased estimates. In particular, models that pool deaths on a common time scale are correctly specified only if the death intensities from different originating states are proportional on that time scale, up to a factor depending on the arrival time, a condition that clock reset after intermediate events can violate. While \autoref{sec.4} explains how these biases arise, their direction and magnitude are tied to our process-based, multi-state data-generating mechanism and the hazard shapes considered, and need not generalize. Furthermore, our evaluation assumed that the time scales of the multi-state model, from which the intermediate-event histories for the CTD models were also simulated, were correctly specified; the robustness of the estimators under time-scale misspecification warrants further investigation.

Finally, our work focuses on population-level treatment effect estimation with only two baseline prognostic factors. For a clinical decision support tool, additional covariates would typically be incorporated, and the target of inference might shift towards individual-level effects, introducing a further layer of uncertainty \citep{bender2025hazards}; in that setting, random-effects models or Bayesian methods could be considered \citep{gerstung2017precision, xu2016bayesian}.

\begin{acknowledgement}
We are grateful to Dr Michal Abrahamowicz and Dr Jan Beyersmann for stimulating discussions and valuable suggestions that helped shape this work, and to the anonymous reviewer for a careful reading of the manuscript and many insightful comments that substantially improved it.

For the original submission, ChatGPT (GPT-5.2, OpenAI) was used for language editing and LaTeX formatting. For the revised version, Claude (Claude Opus 5.5, Anthropic) was used for language editing, LaTeX formatting, checking notation and internal consistency, discussing and revising the authors' draft of Section 4, and updating the existing simulation code for the additional analyses in Section 6. Neither tool was used for study design, method development, or the interpretation of results, and neither influenced the conclusions. The authors reviewed all AI-assisted output and take full responsibility for the content of this article.

\end{acknowledgement}
\vspace*{1pc}

\noindent {\bf{Conflict of Interest}}
\noindent {\it{The authors have declared no conflict of interest.}}

\clearpage
\FloatBarrier
\section*{Appendix A}
\label{app.a}
\setcounter{figure}{0}
\setcounter{table}{0}
\renewcommand{\thefigure}{A.\arabic{figure}}
\renewcommand{\thetable}{A.\arabic{table}}
\renewcommand{\theHtable}{A.\arabic{table}}    
\renewcommand{\theHfigure}{A.\arabic{figure}}
\begin{figure}[!htbp]
    \centering
    \captionsetup{font=small}
    \begin{minipage}{\linewidth}
        \centering
        \includegraphics[width=1\linewidth]{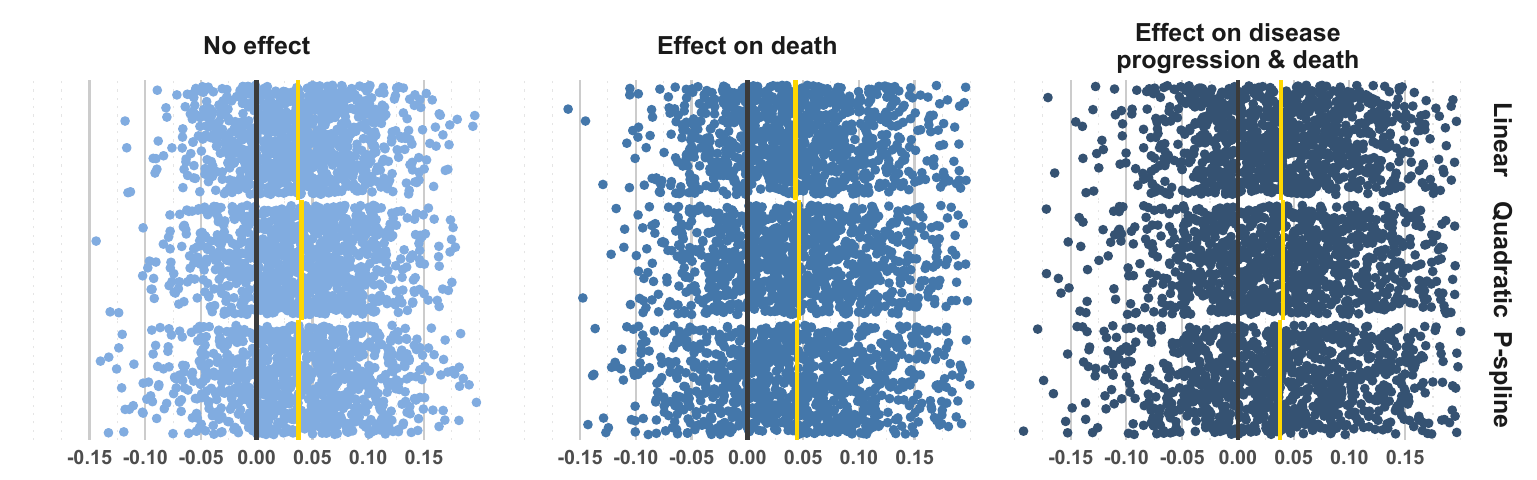}\\[-0.3em]
        \small (a) Semi-Markov
    \end{minipage}

    \vspace{0.8em}

    \begin{minipage}{\linewidth}
        \centering
        \includegraphics[width=1\linewidth]{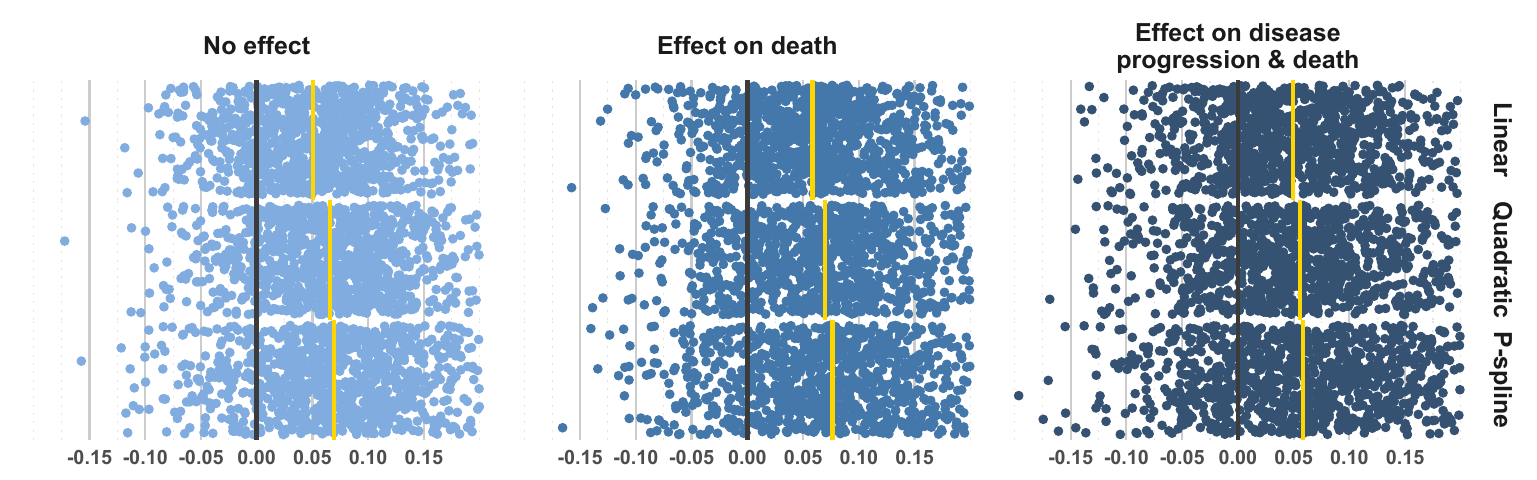}\\[-0.3em]
        \small (b) State arrival extended semi-Markov
    \end{minipage}

    \caption[Impact of functional form of entry time in time-dependent Cox model on estimates (transplantation planned at month 3)]%
    {Sensitivity analysis for the time-dependent Cox model under Weibull death transition intensities: relative bias in the estimated 3-year mortality risk ratio (RR; reference: chemotherapy only) for the strategy \(I_{\mathrm{int}}\) with planned allo-HSCT at \(\tau_{\mathrm{int}}=3\) months since study entry. Points show \((\widehat{\mathrm{RR}}-\mathrm{RR}_{\mathrm{true}})/\mathrm{RR}_{\mathrm{true}}\) across the 1,000 simulation replicates for the time-dependent Cox model under the semi-Markov (a) and state arrival extended semi-Markov (b) data-generating processes. The black vertical line marks zero bias, and the yellow vertical line indicates the mean bias. Results are shown for the three effect settings (no effect; effect on death; effect on disease progression and death) and for alternative specifications of the arrival-time terms in the time-dependent Cox model (linear, quadratic, and penalized spline with 6 degrees of freedom). The linear specification is the one shown in \autoref{fig_2}.}
    \label{fig_3}
\end{figure}

\begin{figure}[htb]
    \centering
    \captionsetup{font=small}
    \includegraphics[width=\linewidth]
    {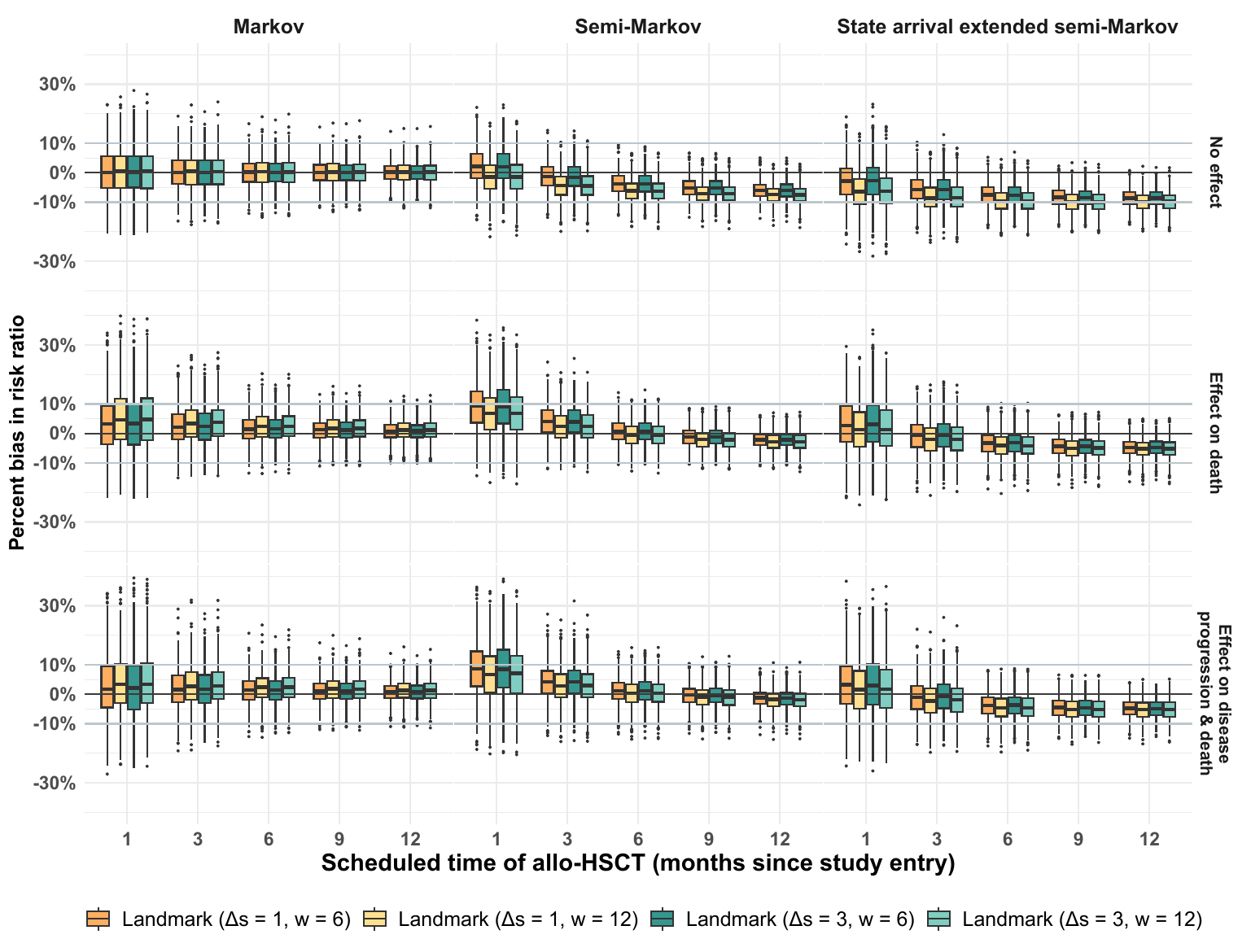}
    \caption{Sensitivity analysis for the landmark supermodel under Weibull death transition intensities: percent bias in the estimated 3-year mortality risk ratio (RR; reference: chemotherapy only), illustrating the impact of landmark supermodel tuning parameters. The x-axis indexes intervention strategies \(I_{\mathrm{int}}\) by planned allo-HSCT time \(\tau_{\mathrm{int}}\) (months since study entry), with one box plot per \(\tau_{\mathrm{int}}\); the y-axis is \(\{(\widehat{\mathrm{RR}}-\mathrm{RR}_{\mathrm{true}})/\mathrm{RR}_{\mathrm{true}}\}\times 100\%\). The specification \(\Delta s=1\), \(w=6\) is the one shown in \autoref{fig_2}.}
    \label{fig_4}
\end{figure}

\clearpage
\phantomsection
\section*{Appendix B}
\label{app.b}
\setcounter{figure}{0}
\setcounter{table}{0}
\renewcommand{\thefigure}{B.\arabic{figure}}
\renewcommand{\thetable}{B.\arabic{table}}
\renewcommand{\theHtable}{B.\arabic{table}}    
\renewcommand{\theHfigure}{B.\arabic{figure}}

\begin{table}[!ht]
\centering
\begin{minipage}{0.6\linewidth}
\centering
\caption{Transition intensity specifications across scenarios with a log-logistic death hazard}
\label{table_b1}

\begin{threeparttable}
\scriptsize
\renewcommand{\arraystretch}{1.15}
\setlength{\tabcolsep}{4pt}

\begin{tabular}{c c c c}
\toprule
& \textbf{Markov} & \textbf{Semi-Markov} &
  \textbf{\begin{tabular}[c]{@{}c@{}}State arrival extended\\ semi-Markov\end{tabular}} \\
\midrule
\multicolumn{4}{c}{\textbf{No treatment effect}}\\
\midrule
$\lambda_{12}(\cdot)$ & $h_{12}(t)$ & $h_{12}(t)$ & $h_{12}(t)$ \\
$\lambda_{13}(\cdot)$ & $h_{13}(t)$ & $h_{13}(t)$ & $h_{13}(t)$ \\
$\lambda_{15}(\cdot)$ & $h_{15}(t)$ & $h_{15}(t)$ & $h_{15}(t)$ \\
$\lambda_{24}(\cdot)$ & $2\cdot h_{13}(t)$ & $2\cdot h_{13}(t-s)$ & $\rho(s)\cdot h_{13}(t-s)$ \\
$\lambda_{56}(\cdot)$ & $h_{12}(t)$ & $h_{12}(t-s)$ & $h_{12}(t-s)$ \\
$\lambda_{57}(\cdot)$ & $h_{13}(t)$ & $h_{13}(t-s)$ & $h_{13}(t-s)$ \\
$\lambda_{68}(\cdot)$ & $2\cdot h_{13}(t)$ & $2\cdot h_{13}(t-s)$ & $\rho(s)\cdot h_{13}(t-s)$ \\
\midrule
\multicolumn{4}{c}{\textbf{Treatment effect on death}}\\
\midrule
$\lambda_{12}(\cdot)$ & $h_{12}(t)$ & $h_{12}(t)$ & $h_{12}(t)$ \\
$\lambda_{13}(\cdot)$ & $h_{13}(t)$ & $h_{13}(t)$ & $h_{13}(t)$ \\
$\lambda_{15}(\cdot)$ & $h_{15}(t)$ & $h_{15}(t)$ & $h_{15}(t)$ \\
$\lambda_{24}(\cdot)$ & $2\cdot h_{13}(t)$ & $2\cdot h_{13}(t-s)$ & $\rho(s)\cdot h_{13}(t-s)$ \\
$\lambda_{56}(\cdot)$ & $h_{12}(t)$ & $h_{12}(t-s)$ & $h_{12}(t-s)$ \\
$\lambda_{57}(\cdot)$ & $0.5\cdot h_{13}(t)$ & $0.5\cdot h_{13}(t-s)$ & $0.5\cdot h_{13}(t-s)$ \\
$\lambda_{68}(\cdot)$ & $2\cdot 0.5\cdot h_{13}(t)$ & $2\cdot 0.5\cdot h_{13}(t-s)$ & $\rho(s)\cdot 0.5\cdot h_{13}(t-s)$ \\
\midrule
\multicolumn{4}{c}{\textbf{Treatment effect on disease progression and death}}\\
\midrule
$\lambda_{12}(\cdot)$ & $h_{12}(t)$ & $h_{12}(t)$ & $h_{12}(t)$ \\
$\lambda_{13}(\cdot)$ & $h_{13}(t)$ & $h_{13}(t)$ & $h_{13}(t)$ \\
$\lambda_{15}(\cdot)$ & $h_{15}(t)$ & $h_{15}(t)$ & $h_{15}(t)$ \\
$\lambda_{24}(\cdot)$ & $2\cdot h_{13}(t)$ & $2\cdot h_{13}(t-s)$ & $\rho(s)\cdot h_{13}(t-s)$ \\
$\lambda_{56}(\cdot)$ & $0.5\cdot h_{12}(t)$ & $0.5\cdot h_{12}(t-s)$ & $0.5\cdot h_{12}(t-s)$ \\
$\lambda_{57}(\cdot)$ & $0.5\cdot h_{13}(t)$ & $0.5\cdot h_{13}(t-s)$ & $0.5\cdot h_{13}(t-s)$ \\
$\lambda_{68}(\cdot)$ & $2\cdot 0.5\cdot h_{13}(t)$ & $2\cdot 0.5\cdot h_{13}(t-s)$ & $\rho(s)\cdot 0.5\cdot h_{13}(t-s)$ \\
\bottomrule
\end{tabular}

\vspace{4pt}
\renewcommand{\TPTnoteSettings}{\scriptsize\raggedright}
\begin{tablenotes}[flushleft]
\item[1] \(h_{12}(t)=0.12 \cdot 0.55 \cdot t^{0.55-1},\; h_{13}(t)=\bigl(0.012 \cdot 1.3 \cdot t^{1.3-1}\bigr)\bigl(1+0.012 \cdot t^{1.3}\bigr)^{-1}\), and \(h_{15}(t)=0.02 \cdot 1.5 \cdot t^{1.5-1}.\)
\item[2] \(s\) is the most recent intermediate-event time before \(t\), or \(0\) if neither has occurred.
\item[3] \(\rho(s)=1+\dfrac{1}{1+\exp(s-6)}.\)
\item[4] Two baseline binary prognostic factors define four equally sized subgroups, with groups 1--4 corresponding to $(Z_1,Z_2)=(0,0),(1,0),(0,1),(1,1)$; group 1 is shown. Weibull scale parameter for $h_{12}$: 0.12,\,\,0.16,\,\,0.12,\,\,0.16; log-logistic inverse-scale parameter for $h_{13}$: 0.012,\,\,0.0125,\,\,0.013,\,\,0.0135.
\item[5] \(t\) and \(s\) are measured in months since study entry.
\end{tablenotes}
\end{threeparttable}
\end{minipage}
\end{table}

\begin{figure}[!htbp]
    \centering
    \includegraphics[width=\linewidth]{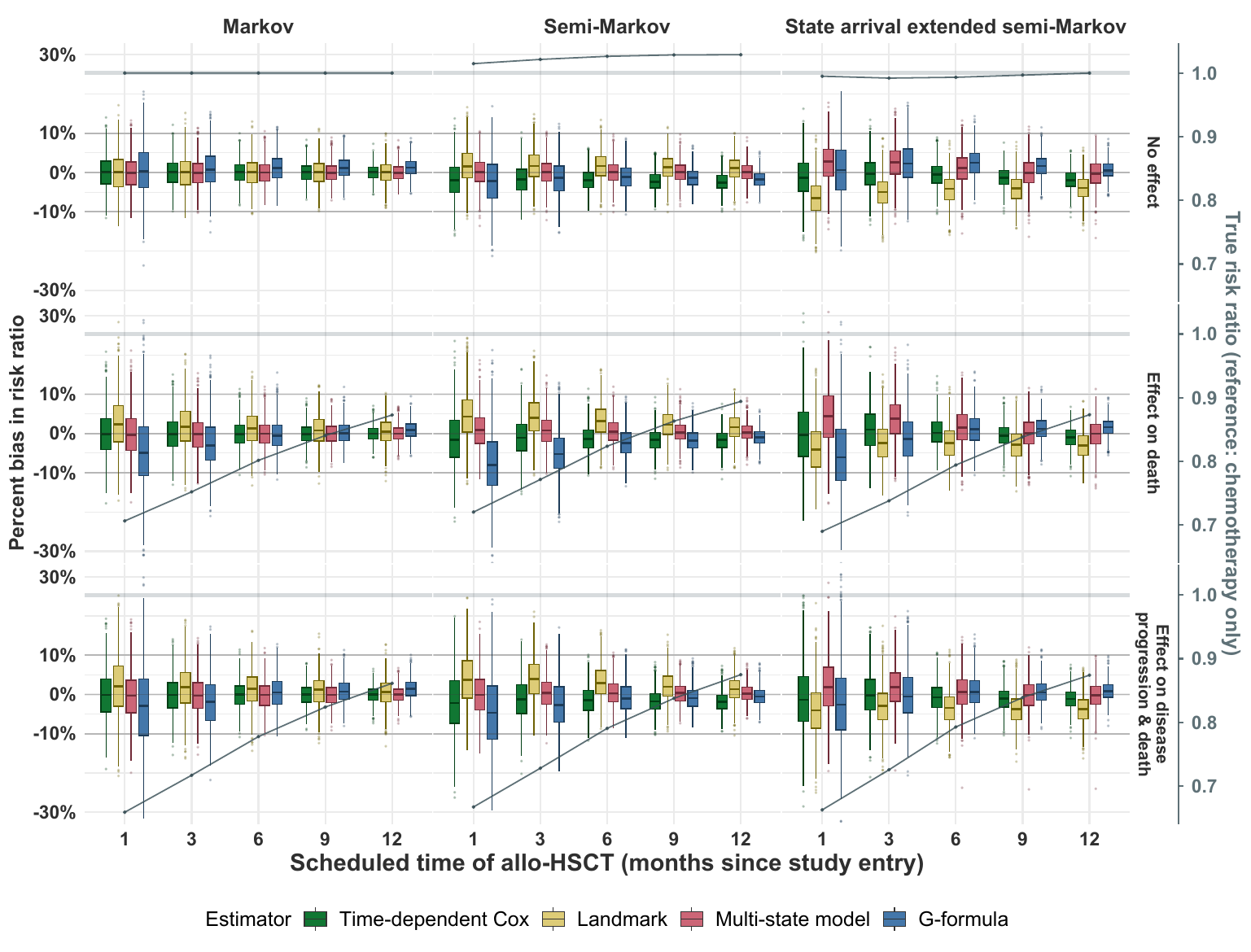}
    \caption[Bias in risk ratio for 3-year mortality (log-logistic distribution)]{Main simulation results under log-logistic death transition intensities: percent bias of the four estimators in the estimated 3-year mortality risk ratio (RR; reference: chemotherapy only). The x-axis indexes intervention strategies \(I_{\mathrm{int}}\) by planned allo-HSCT time \(\tau_{\mathrm{int}}\) (months since study entry), with one box plot per \(\tau_{\mathrm{int}}\); the y-axis is \(\{(\widehat{\mathrm{RR}}-\mathrm{RR}_{\mathrm{true}})/\mathrm{RR}_{\mathrm{true}}\}\times 100\%\). The time-dependent Cox model includes linear terms in the times of allo-HSCT and disease progression in the semi-Markov and state arrival extended semi-Markov settings. The landmark supermodel uses \(\Delta s=1\), \(w=6\), quadratic \(\beta(s)\), and cubic \(g(s)\). The g-formula uses the pooled logistic models of \autoref{sec.2.3} for disease
    progression and death, with a linear term in time, under all three specifications. Solid curves show \(\mathrm{RR}_{\mathrm{true}}\) from the transition intensity matrix. The light horizontal band marks $\mathrm{RR}_{\mathrm{true}} = 1$ (right axis). In the no-effect setting, allo-HSCT has no multiplicative effect on the transition intensities (\autoref{table_b1}); under the semi-Markov and state arrival extended semi-Markov specifications, however, post-transplant intensities restart on the time since allo-HSCT, \textit{e.g.}\ $h_{13}(t-s)$, so that $\mathrm{RR}_{\mathrm{true}}$ need not equal 1.}
   \label{fig_b1}
\end{figure}

\begin{table}[!htbp]
\renewcommand{\arraystretch}{1.20}
\setlength{\tabcolsep}{3.5pt}
\centering
\caption[Simulation performance measures]{Performance measures of risk ratio estimates for 3-year mortality across simulation scenarios under log-logistic death transition intensities at $\tau_{\mathrm{int}}$ = 3 months}
\label{table_b2}

\medskip
\noindent (A) Bias, empirical standard error (EmpSE), and mean squared error (MSE)

\resizebox{\textwidth}{!}{%
\begin{tabular}{l
                *{4}{>{\centering\arraybackslash}m{1.25cm}
                      >{\centering\arraybackslash}m{1.25cm}
                      >{\centering\arraybackslash}m{1.25cm}}}
\toprule
& \multicolumn{3}{c}{\textbf{Time-dependent Cox}}
& \multicolumn{3}{c}{\textbf{Landmark}}
& \multicolumn{3}{c}{\textbf{Multi-state model}}
& \multicolumn{3}{c}{\textbf{G-formula}} \\
\cmidrule(lr){2-4}\cmidrule(lr){5-7}\cmidrule(lr){8-10}\cmidrule(lr){11-13}
& \textbf{Bias} & \textbf{EmpSE} & \textbf{MSE}
& \textbf{Bias} & \textbf{EmpSE} & \textbf{MSE}
& \textbf{Bias} & \textbf{EmpSE} & \textbf{MSE}
& \textbf{Bias} & \textbf{EmpSE} & \textbf{MSE} \\
\midrule

\textbf{No effect} \\
\quad Markov &
0.0001 & 0.0367 & 0.0013 &
0.0001 & 0.0429 & 0.0018 &
-0.0010 & 0.0374 & 0.0014 &
0.0089 & 0.0499 & 0.0026 \\
\quad Semi-Markov &
-0.0175 & 0.0374 & 0.0017 &
0.0173 & 0.0403 & 0.0019 &
0.0011 & 0.0327 & 0.0011 &
-0.0144 & 0.0460 & 0.0023 \\
\quad State arrival extended semi-Markov &
-0.0027 & 0.0422 & 0.0018 &
-0.0496 & 0.0412 & 0.0042 &
0.0253 & 0.0450 & 0.0027 &
0.0238 & 0.0526 & 0.0033 \\
\addlinespace[3pt]

\textbf{Effect on death} \\
\quad Markov &
-0.0017 & 0.0462 & 0.0021 &
0.0191 & 0.0560 & 0.0035 &
-0.0025 & 0.0464 & 0.0022 &
-0.0271 & 0.0632 & 0.0047 \\
\quad Semi-Markov &
-0.0101 & 0.0497 & 0.0026 &
0.0430 & 0.0522 & 0.0046 &
0.0083 & 0.0400 & 0.0017 &
-0.0503 & 0.0577 & 0.0059 \\
\quad State arrival extended semi-Markov &
0.0101 & 0.0570 & 0.0033 &
-0.0231 & 0.0533 & 0.0034 &
0.0372 & 0.0558 & 0.0045 &
-0.0142 & 0.0657 & 0.0045 \\
\addlinespace[3pt]

\textbf{Effect on disease progression \& death} \\
\quad Markov &
-0.0015 & 0.0472 & 0.0022 &
0.0186 & 0.0560 & 0.0035 &
-0.0025 & 0.0475 & 0.0023 &
-0.0188 & 0.0666 & 0.0048 \\
\quad Semi-Markov &
-0.0118 & 0.0541 & 0.0031 &
0.0400 & 0.0532 & 0.0044 &
0.0039 & 0.0426 & 0.0018 &
-0.0240 & 0.0639 & 0.0046 \\
\quad State arrival extended semi-Markov &
0.0000 & 0.0582 & 0.0034 &
-0.0292 & 0.0532 & 0.0037 &
0.0191 & 0.0549 & 0.0034 &
-0.0013 & 0.0681 & 0.0046 \\

\bottomrule
\end{tabular}%
}

\medskip
\noindent (B) Monte Carlo standard error (MCSE) of bias, EmpSE, and MSE
\resizebox{\textwidth}{!}{%
\begin{tabular}{l
                *{4}{>{\centering\arraybackslash}m{1.25cm}
                      >{\centering\arraybackslash}m{1.25cm}
                      >{\centering\arraybackslash}m{1.25cm}}}
\toprule
& \multicolumn{3}{c}{\textbf{Time-dependent Cox}}
& \multicolumn{3}{c}{\textbf{Landmark}}
& \multicolumn{3}{c}{\textbf{Multi-state model}}
& \multicolumn{3}{c}{\textbf{G-formula}} \\
\cmidrule(lr){2-4}\cmidrule(lr){5-7}\cmidrule(lr){8-10}\cmidrule(lr){11-13}
& \textbf{Bias} & \textbf{EmpSE} & \textbf{MSE}
& \textbf{Bias} & \textbf{EmpSE} & \textbf{MSE}
& \textbf{Bias} & \textbf{EmpSE} & \textbf{MSE}
& \textbf{Bias} & \textbf{EmpSE} & \textbf{MSE} \\

\midrule

\textbf{No effect} \\
\quad Markov &
0.0012 & 0.0008 & 0.0001 &
0.0014 & 0.0010 & 0.0001 &
0.0012 & 0.0008 & 0.0001 &
0.0016 & 0.0011 & 0.0001 \\
\quad Semi-Markov &
0.0012 & 0.0008 & 0.0001 &
0.0013 & 0.0009 & 0.0001 &
0.0010 & 0.0007 & 0.0000 &
0.0015 & 0.0010 & 0.0001 \\
\quad State arrival extended semi-Markov &
0.0013 & 0.0009 & 0.0001 &
0.0013 & 0.0009 & 0.0001 &
0.0014 & 0.0010 & 0.0001 &
0.0017 & 0.0012 & 0.0001 \\
\addlinespace[3pt]

\textbf{Effect on death} \\
\quad Markov &
0.0015 & 0.0010 & 0.0001 &
0.0018 & 0.0013 & 0.0002 &
0.0015 & 0.0010 & 0.0001 &
0.0020 & 0.0014 & 0.0002 \\
\quad Semi-Markov &
0.0016 & 0.0011 & 0.0001 &
0.0016 & 0.0012 & 0.0002 &
0.0013 & 0.0009 & 0.0001 &
0.0018 & 0.0013 & 0.0002 \\
\quad State arrival extended semi-Markov &
0.0018 & 0.0013 & 0.0002 &
0.0017 & 0.0012 & 0.0001 &
0.0018 & 0.0012 & 0.0002 &
0.0021 & 0.0015 & 0.0002 \\
\addlinespace[3pt]

\textbf{Effect on disease progression \& death} \\
\quad Markov &
0.0015 & 0.0011 & 0.0001 &
0.0018 & 0.0013 & 0.0002 &
0.0015 & 0.0011 & 0.0001 &
0.0021 & 0.0015 & 0.0002 \\
\quad Semi-Markov &
0.0017 & 0.0012 & 0.0001 &
0.0017 & 0.0012 & 0.0002 &
0.0013 & 0.0010 & 0.0001 &
0.0020 & 0.0014 & 0.0002 \\
\quad State arrival extended semi-Markov &
0.0018 & 0.0013 & 0.0002 &
0.0017 & 0.0012 & 0.0002 &
0.0017 & 0.0012 & 0.0002 &
0.0022 & 0.0015 & 0.0002 \\

\bottomrule
\end{tabular}%
}

\vspace{6pt}
\justifying
\footnotesize
\noindent Notes: $\theta$ is the true value, $\hat\theta_i$ is the estimate from replicate $i$ $(i=1,\ldots,n_{\mathrm{sim}})$, and $\bar\theta$ is their mean.
Bias $=\frac{1}{n_{\mathrm{sim}}}\sum(\hat\theta_i-\theta)$; EmpSE $=\sqrt{\frac{1}{n_{\mathrm{sim}}-1}\sum(\hat\theta_i-\bar\theta)^2}$; MSE $=\frac{1}{n_{\mathrm{sim}}}\sum(\hat\theta_i-\theta)^2$.
The Monte Carlo standard errors of Bias, EmpSE, and MSE reflect only uncertainty due to the finite number of simulation replicates; Monte Carlo variability arising from any within-estimator stochastic approximation is not separately quantified \citep{morris2019using}.
\end{table}

\begin{figure}[!htbp]
    \centering

    \begin{minipage}{\linewidth}
        \centering
        \includegraphics[width=1\linewidth]{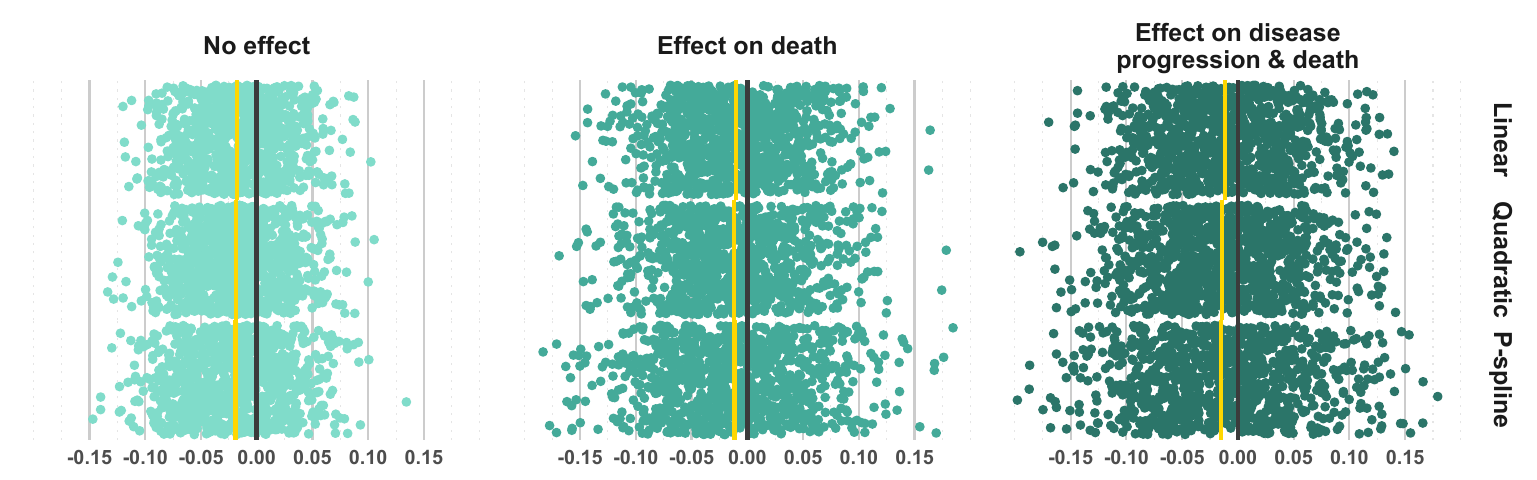}\\[-0.3em]
        \small (a) Semi-Markov
    \end{minipage}

    \vspace{0.8em}

    \begin{minipage}{\linewidth}
        \centering
        \includegraphics[width=1\linewidth]{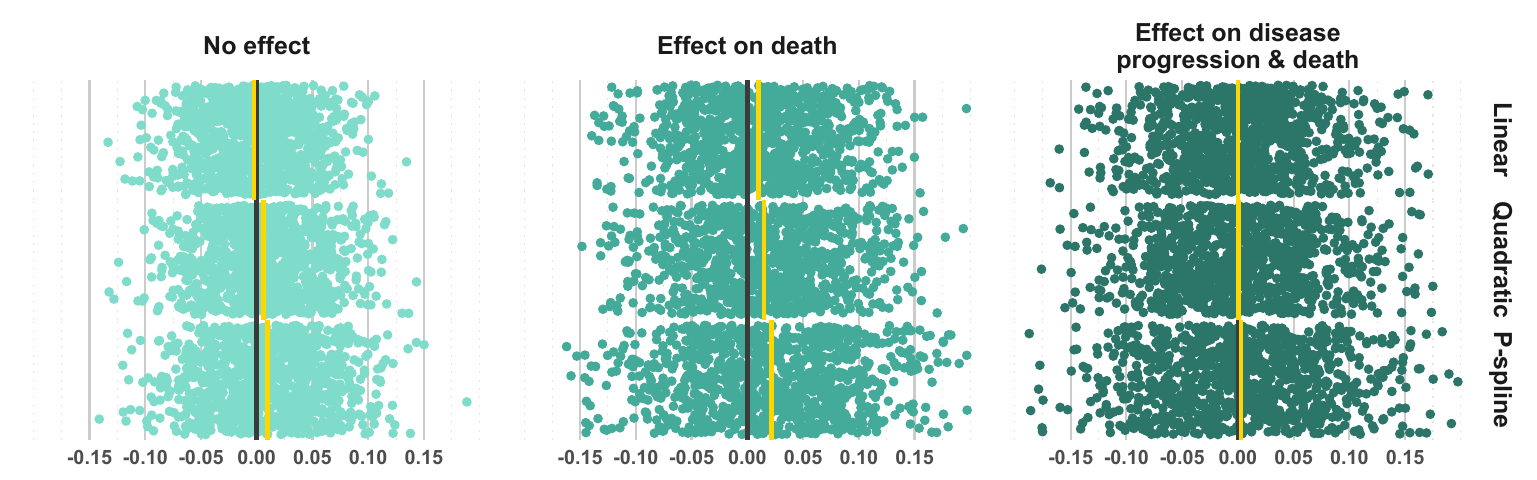}\\[-0.3em]
        \small (b) State arrival extended semi-Markov
    \end{minipage}

    \caption[Impact of functional form of entry time in time-dependent Cox model on estimates (transplantation planned at month 3)]%
    {Sensitivity analysis for the time-dependent Cox model under log-logistic death transition intensities: relative bias in the estimated 3-year mortality risk ratio (RR; reference: chemotherapy only) for the strategy \(I_{\mathrm{int}}\) with planned allo-HSCT at \(\tau_{\mathrm{int}}=3\) months since study entry. Points show \((\widehat{\mathrm{RR}}-\mathrm{RR}_{\mathrm{true}})/\mathrm{RR}_{\mathrm{true}}\) across the 1,000 simulation replicates for the time-dependent Cox model under the semi-Markov (a) and state arrival extended semi-Markov (b) data-generating processes. The black vertical line marks zero bias, and the yellow vertical line indicates the mean bias. Results are shown for the three effect settings (no effect; effect on death; effect on disease progression and death) and for alternative specifications of the arrival-time terms in the time-dependent Cox model (linear, quadratic, and penalized spline with 6 degrees of freedom). The linear specification is the one shown in \autoref{fig_b1}.}
    \label{fig_b2}
\end{figure}

\begin{figure}[htb]
    \centering
    \includegraphics[width=\linewidth]
    {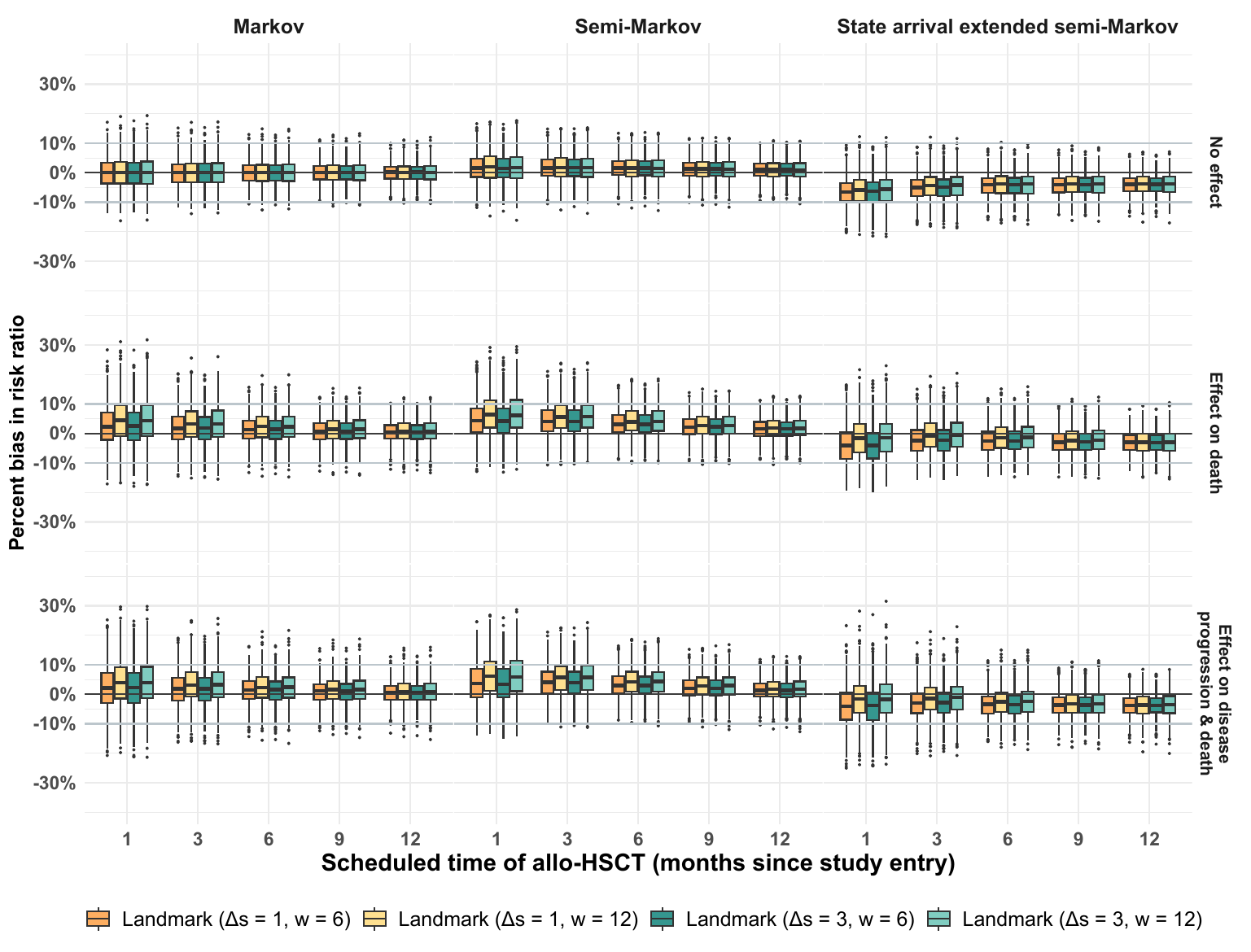}
    \caption{Sensitivity analysis for the landmark supermodel under log-logistic death transition intensities: percent bias in the estimated 3-year mortality risk ratio (RR; reference: chemotherapy only), illustrating the impact of landmark supermodel tuning parameters. The x-axis indexes intervention strategies \(I_{\mathrm{int}}\) by planned allo-HSCT time \(\tau_{\mathrm{int}}\) (months since study entry), with one box plot per \(\tau_{\mathrm{int}}\); the y-axis is \(\{(\widehat{\mathrm{RR}}-\mathrm{RR}_{\mathrm{true}})/\mathrm{RR}_{\mathrm{true}}\}\times 100\%\). The specification \(\Delta s=1\), \(w=6\) is the one shown in \autoref{fig_b1}.}
    \label{fig_b3}
\end{figure}

\clearpage
\FloatBarrier

\bibliographystyle{plainnat_italetal}
\bibliography{Bibliography}

\phantom{aaaa}
\end{document}